\documentclass[10pt,twocolumn]{IEEEtran}
  
\usepackage[utf8]{inputenc}
\usepackage{amssymb,amsmath,amsfonts,amsthm,bm,bbm}
\usepackage{mathrsfs}
\usepackage{graphicx}
\usepackage{cite}
\usepackage{color}
\usepackage{lipsum}
\usepackage[bookmarks]{hyperref}
\usepackage{booktabs}
\usepackage[mathscr]{euscript}
\usepackage[scr]{rsfso}
\usepackage{tablefootnote}
\usepackage{wrapfig}
\usepackage{listings}
\usepackage{array,threeparttable}
\usepackage{makecell}
\usepackage{relsize}
\usepackage{stackengine} 
\usepackage{enumitem} 
\usepackage{enumitem} 

\newcolumntype{M}[1]{>{\centering\arraybackslash}m{#1}}
\def\BibTeX{{\rm B\kern-.05em{\sc i\kern-.025em b}\kern-.08em
    T\kern-.1667em\lower.7ex\hbox{E}\kern-.125emX}}

\usepackage[font=footnotesize]{caption}
\usepackage{float}
\usepackage[list=true]{subcaption}
\usepackage{multirow}
\usepackage[export]{adjustbox}

\DeclareSymbolFont{extraitalic}      {U}{zavm}{m}{it}
\DeclareMathSymbol{\mSampi}{\mathord}{extraitalic}{165}
\DeclareMathSymbol{\msampi}{\mathord}{extraitalic}{166}

\newcolumntype{M}[1]{>{\centering\arraybackslash}m{#1}}
\def\BibTeX{{\rm B\kern-.05em{\sc i\kern-.025em b}\kern-.08em
    T\kern-.1667em\lower.7ex\hbox{E}\kern-.125emX}}
\def\BibTeX{{\rm B\kern-.05em{\sc i\kern-.025em b}\kern-.08em
    T\kern-.1667em\lower.7ex\hbox{E}\kern-.125emX}}
\makeatletter
  \def\hrulefill{\leavevmode\leaders\hrule height 1pt\hfill\kern\z@}
\makeatother
\newcommand{\Laplace}{\mathscr{L}}

\definecolor{gris245}{RGB}{245,245,245}
\definecolor{olive}{RGB}{50,140,50}
\definecolor{brun}{RGB}{175,100,80}
\usepackage{fancybox}

\hypersetup{
  unicode=false,
  colorlinks=true,       
  linkcolor=black,       
  citecolor=black,       
  filecolor=black,       
  urlcolor=black,
  bookmarksdepth=4
}

\newtheorem*{theorem}{Theorem}

\newtheorem*{corollary}{Corollary}
\newtheorem*{definition}{Definition}

\begin{document}


\title{A Gamma-Mixture Solution for Definite Noncentral Complex Gaussian Quadratic Forms}




\author{Fernando D. Almeida García, \textit{Senior Member, IEEE}, Carlos~R.~Nogueira~da~Silva, Felipe A. P. de Figueiredo,\\ Rausley A. A. de Souza, \textit{Senior Member, IEEE}, Michel Daoud Yacoub, \textit{Member, IEEE},\\ and José Cândido Silveira Santos Filho, \textit{Member, IEEE}

\thanks{
This work was funded by the Brasil 6G Project with support from RNP/MCTI (Grant 01245.010604/2020-14), the xGMobile Project (XGM-AFCCT-2025-8-1-1 and XGM-AFCCT-2024-9-1-1) with resources from EMBRAPII/MCTI (Grant 052/2023 PPI IoT/Manufatura 4.0), CNPq (302085/2025-4, 306199/2025-4), FAPEMIG (APQ-03162-24, PPE-00124-23, RED-00194-23), and FINEP (nº 1060/2 contract 01.25.0883.00). We also acknowledge the support of ABRINT, MODIRUM$|$GESPI, NIC.br, Ericsson Brasil, and Volkswagen Truck \& Bus Ltda.

F. D. A. García, F. A. P. de Figueiredo, and R. A. A. de Souza are with the Wireless and Artificial Intelligence (WAI) laboratory, National Institute of Telecommunications (INATEL), Santa Rita do Sapucaí, MG, 37536-001, Brazil (e-mail: $\rm fernando.almeida@ \rm inatel.br$; $\rm felipe.figueiredo@ \rm inatel.br$;
$\rm rausley@ \rm inatel.br$). C. R. Nogueira da Silva is with the Department of Electrical Engineering, Federal University of Triângulo Mineiro, Uberaba 38064-200, Brazil (e-mail: $ \rm carlos.nogueira@ \rm uftm.edu.br$). Michel Daoud Yacoub and José Cândido Silveira Santos Filho are with the Department of Communications, School of Electrical and Computer Engineering, State University of Campinas, Campinas, SP, 13083-852, Brazil (e-mail: $ \rm mdyacoub@ \rm unicamp.br$; $ \rm jcssf@ \rm unicamp.br$)
}
}


\maketitle

\begin{abstract}
Finding \emph{exact} solutions for the statistics of definite noncentral complex Gaussian quadratic forms (CGQFs) has long been a challenging problem.
Among existing exact approaches, Raphaeli's formulation~\cite{Raphaeli96} remains the most computationally efficient. 
However, its analytical tractability is hindered by the dependence of its series coefficients on the evaluation variable, which confines their statistics to the probability density function (PDF) and cumulative distribution function (CDF). This paper overcomes this limitation by deriving novel Gamma-mixture representations for the PDF and CDF of positive- and negative-definite noncentral CGQFs. The resulting structure broadens Raphaeli's statistical framework beyond the PDF and CDF to a wider class of CGQF functionals and a substantial range of practical applications.
In particular, expectations of integrable functions of definite noncentral CGQFs are expressed as weighted sums of the corresponding expectations over Gamma random variables, enabling the direct derivation of moments, transforms, and performance measures.
Moreover, the proposed formulation reduces the single-point complexity from \(O\!\left(n^2K+nK^2\right)\) for Raphaeli's method to \(O\!\left(nK+K^2\right)\), where \(n\) denotes the quadratic-form dimension and \(K\) the truncation order.
The proposed Gamma-mixture structure also enables closed-form approximations that, by construction, remain valid probability distributions for any truncation order. In addition, uniform PDF and CDF truncation-error bounds are derived, allowing prescribed-accuracy truncation. 
As a sample application, we derive the first reported exact average bit error rate expression for maximal-ratio combining receivers with an arbitrary number of branches under an arbitrary positive-definite Rician correlation structure.

\end{abstract}

\begin{IEEEkeywords}
Definite noncentral complex Gaussian quadratic forms, Gamma-mixtures, probability density function, cumulative distribution function, performance measures.
\end{IEEEkeywords}

\section{Introduction}
\label{sec: Introduction}

\IEEEPARstart{G}{aussian} quadratic forms (GQFs) arise in a broad range of wireless communications and signal-processing applications, including optimum combining~\cite{Lao2003}, noncoherent modulation~\cite{Raphaeli1996Noncoherent,Divsalar1990,Huang1991}, diversity techniques~\cite{Rao2001,Garzon24,Garcia22alpha_mu,Garcia_Mixed_Sums}, beamforming~\cite{Kim2013,Sharif2005}, multibeam transmission strategies~\cite{Schlegel1996}, orthogonal space--time block coding~\cite{Ropokis2008}, relay-assisted communications~\cite{HavaryNassab2010,Zhao2004,Zhao2005,ShahLi2005}, analysis of variance~\cite{Johnson1994}, power-spectrum estimation~\cite{Grenander1959}, differential detection~\cite{Pauli2008}, matched-field processing~\cite{Gall2014}, cooperative communications~\cite{Wornell2001}, and adaptive filtering~\cite{AlNaffouri2010,Bershad1987,Bershad1986}.
GQFs can be classified according to the nature of the underlying Gaussian random variables, the definiteness of the associated quadratic-form matrix, and the presence or absence of noncentrality. Accordingly, they may be real or complex; positive definite, negative definite, or indefinite; and central or noncentral~\cite{Provost1992}.\footnote{ Formulations developed for indefinite GQFs may encompass the positive- and negative-definite cases as special cases when all eigenvalues have the same sign~\cite{Provost1992}.} As the model incorporates a broader combination of these characteristics, deriving its fundamental distributional quantities---particularly the probability density function (PDF) and cumulative distribution function (CDF)---becomes more challenging. In this paper, we focus on positive- and negative-definite noncentral complex Gaussian quadratic forms (CGQFs).\footnote{For convenience, unless otherwise stated, the term ``definite'' is used collectively to refer to both positive-definite and negative-definite cases. For instance, ``definite noncentral CGQFs'' refers to positive- and negative-definite noncentral CGQFs.
}


\subsection{Related Work on Gaussian Quadratic Forms}
\label{subsec:related_work}

\subsubsection{Exact Approaches}

The distribution of GQFs has been studied extensively since the mid-twentieth century. Early contributions focused primarily on quadratic forms (QFs) in real Gaussian variables. Robbins~\cite{Robbins1948} and Gurland~\cite{Gurland1953,Gurland1955,Gurland1956} developed distributional representations and characteristic-function inversion methods for definite and indefinite quadratic forms, whereas Shah~\cite{Shah1963} extended these analyses to noncentral normal variables.
The study of CGQFs can be traced to Turin~\cite{Turin1960}, who derived the characteristic function of Hermitian quadratic forms in complex Gaussian variables. Kotz \mbox{\emph{et al.}}~\cite{Kotz67central,Kotz67} later derived series expansions for central and noncentral positive-definite real GQFs using Maclaurin-series, Laguerre-polynomial, and chi-squared-series expansions. 
Tziritas~\cite{Tziritas87} derived series representations for the PDF and CDF of central and noncentral positive-definite GQFs involving real and complex Gaussian variables, whereas Reifler \mbox{\emph{et al.}}~\cite{Reifler89} derived a multidimensional Hermite-polynomial series representation for the CDF of noncentral positive-definite real GQFs. For indefinite CGQFs, Biyari and Lindsey~\cite{Biyari93} derived exact series expressions for a particular class of noncentral Hermitian GQFs. 
Raphaeli~\cite{Raphaeli96} subsequently derived series representations for the PDF and CDF of noncentral indefinite CGQFs using contour integration and residue calculus, providing a computationally efficient and fast-converging exact approach.

\subsubsection{Approximate Approaches}

Approximate solutions were later developed to reduce the computational burden associated with exact series and numerical inversion methods. Saddlepoint approximations were investigated in~\cite{Marsh1998,Kuonen99}. Al-Naffouri \mbox{\emph{et al.}} subsequently proposed a general framework encompassing definite and indefinite, as well as central and noncentral GQFs~\cite{Naffouri16}. Their central-case results are available in closed form, whereas the noncentral case is represented by a one-dimensional integral that can be efficiently evaluated using a saddle-point approximation.
Ramírez-Espinosa \mbox{\emph{et al.}} proposed an alternative approximation for noncentral positive-definite real GQFs based on a linear combination of derivatives of the moment-generating function~\cite{Espinosa19SP}.
Ramírez-Espinosa \mbox{\emph{et al.}} later introduced an auxiliary-variable framework applicable to both definite and indefinite noncentral CGQFs~\cite{Espinosa19}. The resulting PDF and CDF expressions define valid approximate distributions for every value of the accuracy-controlling parameter and converge to the target distribution as this parameter increases.

\subsubsection{Numerical Methods}

Imhof introduced a numerical characteristic-function inversion method for evaluating the distributions of real GQFs~\cite{Imhof1961}. Davies later developed a related inversion procedure with numerical error control~\cite{Davies73}, whereas Rice proposed numerical-integration techniques for computing GQF distributions~\cite{Rice1980}. 
These methods avoid the evaluation of complicated series coefficients and can provide accurate pointwise estimates of the PDF or CDF. 
More recently, Issaid \mbox{\emph{et al.}} developed an importance-sampling method for the efficient estimation of rare-event and tail probabilities associated with positive GQFs~\cite{Issaid21}.

\subsection{Drawbacks and Limitations}
\label{subsec:drawbacks}

Despite substantial progress, existing exact and approximate approaches involve important tradeoffs among scope, computational efficiency, convergence behavior, and analytical tractability. Their principal
limitations can be summarized as follows:

\begin{itemize}

\item \emph{Restricted applicability:}
Several formulations are tailored to specific subclasses of GQFs and
therefore do not provide a general treatment of real and complex, central and noncentral, and definite and indefinite cases. For example, the results
in~\cite{Kotz67central,Kotz67,Reifler89} are restricted to central or
noncentral positive-definite real GQFs, while the approximation
in~\cite{Espinosa19SP} is likewise limited to noncentral
positive-definite real GQFs. Tziritas~\cite{Tziritas87}, in turn,
considered central and noncentral positive-definite GQFs involving both real
and complex Gaussian variables. The formulation of Biyari and
Lindsey~\cite{Biyari93} applies only to a particular class of noncentral
Hermitian GQFs. Raphaeli's exact formulation~\cite{Raphaeli96} and the approximate frameworks of Ramírez-Espinosa \mbox{\emph{et al.}}~\cite{Espinosa19} and Al-Naffouri \mbox{\emph{et al.}}~\cite{Naffouri16} represent the broadest class of available approaches, encompassing noncentral indefinite CGQFs.



\item \emph{High computational complexity:}
The cost of exact nested-series methods generally increases with the
quadratic-form dimension, eigenvalue multiplicities, noncentrality
parameters, and required truncation order
\cite{Shah1963,Kotz67central,Kotz67,Biyari93,Raphaeli96,Provost1992}.
In Raphaeli's formulation~\cite{Raphaeli96}, for example, clusters of closely spaced eigenvalues may further slow convergence, thereby increasing the required truncation order and the overall computational burden.
Approximate methods are subject to similar scaling limitations. In~\cite{Espinosa19SP}, the computational cost increases with the quadratic-form dimension and the auxiliary shape parameter, as larger values require higher-order derivatives of the moment-generating function. Likewise, in the formulation of~Ramírez-Espinosa \mbox{\emph{et al.}}~\cite{Espinosa19}, the cost of evaluating the series coefficients grows combinatorially with the quadratic-form dimension and the accuracy-controlling parameters.


\item \emph{Validity of the PDF and CDF:}
For practical implementation, the exact PDF and CDF series must be truncated.
However, the resulting truncated versions are not automatically guaranteed
to preserve all the defining properties of a probability distribution. In
particular, the truncated PDF may become negative or fail to integrate
exactly to unity, while the truncated CDF may take values outside
\([0,1]\), as discussed in~\cite{Espinosa19}. Similar consistency issues may also arise in approximate methods. For example, the approximation
in~\cite{Naffouri16} may yield CDF values greater than one in the right
tail, as observed in~\cite{Espinosa19SP}.


\item \emph{Limited analytical tractability:}
Some exact and approximate formulations are difficult to use in subsequent
analytical derivations. In Raphaeli's method~\cite{Raphaeli96}, the
recursively generated coefficients depend on the evaluation variable,
complicating integration, differentiation, and transformation. The
approximate solution in~\cite{Naffouri16} relies on a one-dimensional
integral and saddlepoint evaluation, so additional expectations generally
require further numerical integration. Likewise, the approximate method
in~\cite{Espinosa19SP} may require high-order derivatives of the
moment-generating function, which must be computed numerically when
closed-form expressions are unavailable. These limitations hinder the
derivation of moments, transforms, and average performance measures.


\end{itemize}

\subsection{Our Contributions}

Motivated by the drawbacks and limitations discussed previously, this paper derives analytically tractable and computationally efficient formulations for the
\emph{exact} characterization of definite noncentral
CGQFs. The principal contributions are summarized as follows:

\begin{enumerate}
    \item We derive novel exact Gamma-mixture representations for the PDF and CDF of positive- and negative-definite noncentral CGQFs. A key feature is that the mixture weights are independent of the evaluation variable, so they can be computed once and reused across multiple points. Unlike Raphaeli's formulation~\cite{Raphaeli96}, whose coefficients depend on the evaluation variable, the proposed structure yields a simpler and more tractable formulation, while improving computational efficiency.

    \item We establish the absolute and uniform convergence of the PDF and CDF series and derive uniform upper bounds on their absolute truncation errors. These bounds enable systematic selection of a sufficient truncation order to guarantee a prescribed accuracy.


     \item We show that expectations of integrable functions of definite noncentral CGQFs can be expressed as a weighted sum of the corresponding expectations over Gamma random variables,
     enabling the direct derivation of moments, transforms, and performance measures. As a sample application, we derive an exact average bit error rate (ABER) expression for square $M$-ary quadrature amplitude modulation ($M$-QAM) in maximal-ratio combining (MRC) receivers over correlated Rician fading channels. To the best of our knowledge, this is the first exact result for an arbitrary number of diversity branches under arbitrary Rician correlation.

     \item We derive the computational complexity orders of the proposed PDF and CDF representations and compare them with those of Raphaeli's method. The analysis shows that the proposed solutions have lower computational complexity, with the advantage becoming particularly pronounced for multipoint evaluations, i.e., when generating complete PDF and CDF curves over their full support.

     \item Capitalizing on the Gamma-mixture structure of the proposed representations, we derive finite-sum closed-form approximations by normalizing the retained mixture weights. Unlike direct truncations of existing exact series, the proposed approximations define valid probability distributions for every truncation order: the approximate PDF remains nonnegative and integrates to unity, while the corresponding CDF preserves the correct support, range, and limiting behavior.

    \item We validate the proposed analytical representations through Monte Carlo simulations. The numerical results show that the proposed method outperforms Raphaeli's formulation~\cite{Raphaeli96} in terms of computational complexity and execution time.
    For high-dimensional quadratic forms and a prescribed accuracy level, our method also surpasses the approximation proposed by Ramírez-Espinosa \mbox{\emph{et al.}}~\cite{Espinosa19} in terms of execution time.
    The approximate results of Al-Naffouri \mbox{\emph{et al.}}~\cite{Naffouri16} are also included as accuracy benchmarks.

\end{enumerate}

\subsection{Notation}

Throughout the paper, \(\Pr\{\cdot\}\) denotes probability;
\(f_{(\cdot)}(\cdot)\), PDF;
\(F_{(\cdot)}(\cdot)\), CDF;
\(\mathbb{E}\{\cdot\}\), expectation;
\(\Re\{\cdot\}\), the real part;
\(\Gamma(\cdot)\), the gamma function;
\(\Upsilon(\cdot,\cdot)\), the lower incomplete gamma function; \(I_{(\cdot)}(\cdot,\cdot)\), the regularized incomplete beta function;
\(\Gamma(\cdot,\cdot)\), the upper incomplete gamma function; \(\mathbb{N}\), the set of natural numbers;
\(\mathbb{N}_0\), the set of natural numbers including zero;
\(\mathbb{R}\), the set of real numbers;
\(\mathbb{C}\), the set of complex numbers;
\(\operatorname{u}(\cdot)\), the unit-step function;
\((\cdot)^\dagger\), the conjugate transpose;
\(\operatorname{sgn}(\cdot)\), the sign function;
\(\operatorname{card}(\mathcal{A})\), the cardinality of the set
\(\mathcal{A}\);
\(\mathcal{CN}_n(\boldsymbol{a},\boldsymbol{B})\), an
\(n\)-dimensional complex Gaussian distribution with mean vector
\(\boldsymbol{a}\) and covariance matrix \(\boldsymbol{B}\);
\(\chi_a^2(b)\), a noncentral chi-square distribution with \(a\)
degrees of freedom and noncentrality parameter \(b\);
\(\Laplace\{\cdot\}(s)\), the Laplace transform evaluated at \(s\);
\(O(\cdot)\), the asymptotic order of computational complexity;
and \(\sim\), ``distributed as.''
Moreover, \(\max_{i}\{a_i\}\) and \(\min_{i}\{a_i\}\) denote the maximum and minimum elements of the set \(\{a_i\}_{i=1}^{n}\), respectively, while \(\sup_{x\in\mathcal{A}} g(x)\) and \(\inf_{x\in\mathcal{A}} g(x)\) denote the supremum and infimum of \(g(x)\) over \(\mathcal{A}\), respectively.


\section{Preliminaries: Definite Noncentral Complex Gaussian Quadratic Forms}
\label{sec:preliminaries}

This section introduces definite noncentral CGQFs and establishes the
mathematical foundation underpinning the subsequent analysis.





\begin{definition}[CGQF]
Let $\mathbf{v}\sim\mathcal{CN}_n(\bar{\mathbf{v}},\mathbf{L})$ be a complex Gaussian vector with mean $\bar{\mathbf{v}}\in\mathbb{C}^n$ and full-rank Hermitian covariance 
matrix $\mathbf{L}\in\mathbb{C}^{n\times n}$. For a Hermitian matrix 
$\mathbf{A}\in\mathbb{C}^{n\times n}$, the quadratic form
\begin{equation}
    \label{eq: Q def}
    Q = \mathbf{v}^\dagger \mathbf{A}\mathbf{v}
\end{equation}
is called a \emph{noncentral complex Gaussian quadratic form (CGQF)}. 
It is said to be \emph{definite} if all eigenvalues of $\mathbf{A}$ are strictly positive or strictly negative, and \emph{semi-definite} if all eigenvalues are non-negative or non-positive, with at least one eigenvalue equal to zero.
\end{definition}

It was shown by Tziritas~\cite{Tziritas87} and Khuri~\cite{Khuri99} that the noncentral CGQF in \eqref{eq: Q def} can be expressed as
\begin{equation}
\label{eq: Q form}
Q = \sum_{i=1}^{n} \eta_i \, Y_i ,
\end{equation}
where $\left\{ Y_i \right\}_{i=1}^n$ are mutually independent random variables with $Y_i \sim \chi^2_{2}\!\left(2\mu_i\right)$, $\eta_i \triangleq \lambda_i/2$,
$\{\lambda_i\}_{i=1}^{n}$ denote the eigenvalues of $\mathbf{L}\mathbf{A}$,
and the parameters $\{\mu_i\}_{i=1}^{n}$ are given by
\begin{equation}
\mu_i \triangleq
\left[
\mathbf{U}^\dagger
\mathbf{C}^{-1}
\bar{\mathbf{v}}
\bar{\mathbf{v}}^\dagger
\left(\mathbf{C}^{-1}\right)^\dagger
\mathbf{U}
\right]_{i,i} \geq 0,
\end{equation}
where $\mathbf{C} \in \mathbb{C}^{n \times n}$ is an invertible lower triangular matrix
with non-negative diagonal entries obtained from the Cholesky factorization
$\mathbf{L} = \mathbf{C}\mathbf{C}^\dagger$, and $\mathbf{U}$ is the unitary matrix
that diagonalizes $\mathbf{C}^\dagger \mathbf{A}\mathbf{C}$.
The representation in~\eqref{eq: Q form} holds without loss of generality for
positive or negative semi-definite noncentral CGQFs. In these cases, one or more
eigenvalues $\{\lambda_i\}$ may be equal to zero, in which case the corresponding
terms in~\eqref{eq: Q form} vanish identically. Consequently, if $r$ eigenvalues
are zero, after re-indexing the nonzero eigenvalues, the effective summation range in~\eqref{eq: Q form} reduces to
$i = 1,\ldots,n-r$.

More recently, García \mbox{\emph{et al.}}~\cite{Garcia24} showed that the
PDF and CDF of sums of independent squared Gaussian random variables with
nonzero deterministic or random means admit Gamma-mixture representations.
Combined with the canonical decomposition of definite noncentral CGQFs
established by Tziritas~\cite{Tziritas87} and Khuri~\cite{Khuri99}, this
result suggests that the distribution of \(Q\) in \eqref{eq: Q form} may
likewise admit a Gamma-mixture characterization. This observation motivates
the analytical developments presented in the next section.

\section{Fundamental Statistics}
\label{sec: Statistics}

This section derives novel exact expressions for the fundamental statistics of definite noncentral CGQFs, analyzes the computational complexity of the proposed formulation, and compares it with Raphaeli's exact method~\cite{Raphaeli96}. Uniform
truncation-error bounds are subsequently established, followed by the
development of closed-form approximations for the practical evaluation of the PDF and CDF.

\subsection{Main Result}

\label{sec: Main Result}
\begin{theorem}
Let \(Q\) be a definite noncentral CGQF, as defined in \eqref{eq: Q def}.
The PDF and CDF of $Q$
admit, respectively, the following Gamma-mixture representations
\begin{align}
    \label{eq: theorem PDF Q}
    f_Q(q) \;=\; & \sum_{k=0}^{\infty} \beta_k \,
    \mathit{f}_{G}\left(\alpha_k,\theta,\left|q\right| \right) \, \operatorname{u}(q\,\eta^{*}), & q\in\mathbb{R} \\ \label{eq: theorem CDF Q}
    \nonumber F_Q(q) \;=\; & \frac{1- \sigma}{2} &\\
     + & \sigma   \sum_{k=0}^{\infty}\beta_k \,
    \mathit{F}_{G}\left(\alpha_k,\theta,\left|q\right|\right) \operatorname{u}(q\,\eta^{*}), & q\in\mathbb{R},
\end{align}
where  $\sigma \triangleq \operatorname{sgn}(\eta^*)$, $\theta \triangleq 2 \left|\eta^{*}\right|$, $\alpha_k \triangleq k + n$, and $\eta^{*}$ is defined as
\begin{align}
\label{eq: eta cases}
\eta^{*} \triangleq
\begin{cases}
\displaystyle \min_{i}\{\eta_i\}, & \text{if } \eta_i > 0\ \forall i \\
\displaystyle \max_{i}\{\eta_i\}, & \text{if } \eta_i < 0\ \forall i .
\end{cases}
\end{align}
Furthermore,
\begin{align}
    \label{eq: PDF G Gamma marginal}
    \mathit{f}_{G}\left(\alpha_k,\theta,x\right) = & \frac{ x^{\alpha_k-1}}{\theta^{\alpha_k} \Gamma \left(\alpha_k\right)} \exp \left(-\frac{x}{\theta}\right), & x \geq 0\\ \label{eq: CDF G Gamma marginal}
    \mathit{F}_{G}\left(\alpha_k,\theta,x\right) = & \frac{ \Upsilon \left(\alpha_k,\frac{x}{\theta}\right)}{\Gamma(\alpha_k)}, & x \geq 0,
\end{align}
denote the PDF and the CDF of a Gamma distribution with shape parameter $\alpha_k$ and scale parameter $\theta$, respectively.
The mixture weights $\left\{\beta_k \right\}_{k=0}^{\infty}$ satisfy $\beta_k \geq 0$ $\forall k$ and
\mbox{$\sum_{k=0}^{\infty} \beta_k = 1$}, and are defined recursively as
\begin{subequations}
\label{eq: Coefficients beta final}
\begin{align}
\label{eq: beta exp 0 final}
\beta_0 =& \left(\prod_{i=1}^{n} r_i \right)
\exp\!\left(-\sum_{i=1}^{n} \mu_i\right), \\
\label{eq: beta exp i final}
\beta_k = &\frac{1}{k} \sum_{m=1}^{k} \beta_{k-m} \sum _{i=1}^n a_i^{m-1} \left(a_i+ r_i m \, \mu _i\right)
\qquad k \geq 1,
\end{align}
\end{subequations}
where $ r_i \triangleq \frac{\eta^*}{\eta_i} = \frac{|\eta^*|}{|\eta_i|}$ and $a_i \triangleq 1-r_i$.



\end{theorem}

\begin{proof}
    See Appendix~\ref{app: proof theorem}.
\end{proof}

The absolute and uniform convergence of the PDF and CDF series in \eqref{eq: theorem PDF Q} and \eqref{eq: theorem CDF Q} is established in Appendix~\ref{app: series convergence}.

\textit{Remark 1:} 
To the best of our knowledge, the Gamma-mixture representations in
\eqref{eq: theorem PDF Q} and \eqref{eq: theorem CDF Q} are novel and
arguably constitute the most efficient exact solutions available for this
class of definite noncentral CGQFs.
Notably, the Gamma-mixture structure is derived directly from the bilateral Laplace transform of the QF, rather than being imposed through a prespecified basis expansion or assumed distributional form, as in~\cite{Kotz67central,Kotz67,Tziritas87,Reifler89}. This
probabilistic structure naturally yields valid closed-form
approximations, as shown in
Section~\ref{sec: closed-form approximations}. Furthermore, as the mixture weights are independent of the evaluation variable, the proposed representation facilitates subsequent analytical derivations, as established in the following corollary.

\begin{corollary}[Expectations of functions of \(Q\)]
\label{cor: expectation Q}
Let \(Q\) be a definite noncentral CGQF, as defined in \eqref{eq: Q def}, and let \(\psi:\mathbb{R}\to\mathbb{C}\) be a Borel-measurable function satisfying \(
\mathbb{E}\!\left\{|\psi(Q)|\right\}<\infty\).
Then,
\begin{equation}
\mathbb{E}\!\left\{\psi(Q)\right\}
=
\sum_{k=0}^{\infty}
\beta_k
\mathbb{E}\!\left\{\psi(\sigma G_k)\right\},
\label{eq: expectation Q gamma mixture}
\end{equation}
where $\mathbb{E}\!\left\{\psi(\sigma G_k)\right\}$ is the expectation of the function $\psi(\sigma G_k)$, and \(G_k\) is a Gamma random variable with shape parameter \(\alpha_k\) and scale parameter \(\theta\).
The parameters \(\beta_k\), \(\sigma\), \(\alpha_k\), and \(\theta\) are defined in the preceding theorem. The series in \eqref{eq: expectation Q gamma mixture} converges absolutely.
\end{corollary}

\begin{proof}
See Appendix~\ref{app: proof expectation Q}.
\end{proof}

\textit{Remark 2:}
The corollary establishes that the expectation of any integrable function of a definite noncentral CGQF can be expressed as weighted sums of expectations of the same function evaluated at univariate Gamma random variables. The absolute convergence of the resulting series ensures that this representation is well defined and permits term-by-term numerical evaluation. For a positive-definite CGQF, $\sigma=1$, so that the component terms reduce to $\mathbb{E}\{\psi(G_k)\}$. For a negative-definite CGQF, $\sigma=-1$, and the sign reversal is incorporated through $\mathbb{E}\{\psi(-G_k)\}$. Therefore, whenever the corresponding expectation involving a Gamma random variable admits a closed-form expression, the associated functional of $Q$ follows directly from \eqref{eq: expectation Q gamma mixture}. This representation thus provides a general mechanism for deriving moments, integral transforms, average error probabilities, detection probabilities, and other average performance measures involving definite noncentral CGQFs.

A particularly useful consequence arises in wireless communication applications. When system performance depends on a definite noncentral CGQF, metrics such as the ABER can be expressed as weighted sums of the corresponding ABER expressions associated with a \emph{single-link} Nakagami-$m$ fading channel, as illustrated in Section~\ref{sec: Sample Application}. This connection follows from the fact that the instantaneous power under Nakagami-$m$ fading is Gamma distributed. 

\subsection{Computational Complexity}
\label{sec: order of complexity}

Here, we analyze the computational complexity of the truncated PDF and
CDF representations in \eqref{eq: theorem PDF Q} and
\eqref{eq: theorem CDF Q}, respectively, and compare their complexity
with that of Raphaeli's method~\cite{Raphaeli96}. The
analysis is carried out by retaining the terms indexed by
\(k=0,\ldots,K\). 
Throughout this analysis, fixed-precision arithmetic operations and
evaluations of standard built-in special functions are assumed to have
constant computational cost.

\subsubsection{Complexity of the Proposed Method}

The component-wise computational costs of the proposed Gamma-mixture representations are summarized as follows.

\begin{itemize}

    \item \textit{Gamma-component evaluation:}
    For a fixed value of \(q\), evaluating either
    \(
        f_G\!\left(\alpha_k,\theta,|q|\right)
    \)
    in \eqref{eq: PDF G Gamma marginal} or
    \(
        F_G\!\left(\alpha_k,\theta,|q|\right)
    \)
    in \eqref{eq: CDF G Gamma marginal} requires a constant number of
    arithmetic operations and one evaluation of a standard Gamma or
    incomplete-Gamma function. Hence, each Gamma component has
    complexity \(O(1)\).
    Since each truncated series contains \(K+1\) components, evaluating
    either the PDF or the CDF, once the mixture weights are available,
    requires \(O(K)\) operations at each value of \(q\).

    \item \textit{Initial mixture weight:}
    Computing \(\beta_0\) requires one product and one sum over
    \(n\) terms. Therefore, its computational complexity is
    \(O(n)\).

    \item \textit{Recursive mixture weights:}
    From \eqref{eq: appA Sm}, the recursion in \eqref{eq: beta exp i final} can be written as
    \begin{align}
        \beta_k
        =
        \frac{1}{k}
        \sum_{m=1}^{k}
        \beta_{k-m}S_m,
        \qquad k\geq1.
        \label{eq: complexity beta Sm}
    \end{align}

    A key computational feature is that \(S_m\) depends on \(m\), but
    not on \(k\). Therefore, the quantities
    \(\{S_m\}_{m=1}^{K}\) can be computed once and subsequently reused
    in the recursion for all mixture weights. Thus, computing all \(K\) quantities
    \(S_m\) has complexity $\mathcal{C}_{S}
        =
        O(nK)$.
    Once \(\{S_m\}_{m=1}^{K}\) are available, evaluating
    \(\beta_k\) through \eqref{eq: complexity beta Sm} requires
    \(O(k)\) operations. Hence, generating all recursive mixture
    weights requires $\sum_{k=1}^{K}O(k)
        =
        O\!\left(K^2\right)$.
    Including the computation of \(\beta_0\), the total complexity
    required to generate the mixture weights is therefore
    \begin{align}
        \mathcal{C}_{\beta}
        &=
        O(n)
        +
        O(nK)
        +
        O(K^2)
       =
        O\!\left(
            nK+K^2
        \right).
        \label{eq: complexity beta total}
    \end{align}

    \item \textit{Overall complexity:}
    After the mixture weights have been generated, evaluating and
    accumulating the \(K+1\) Gamma components requires an additional
    \(O(K)\) operations for either the PDF or the CDF. Therefore, the
    total computational complexities at a single evaluation point
    \(q\) are
    \begin{align}
        \mathcal{C}_{\mathrm{PDF}}
        =
        \mathcal{C}_{\mathrm{CDF}}
        &=
        O\!\left(
            nK+K^2+K
        \right)
        =
        O\!\left(
            nK+K^2
        \right).
        \label{eq: complexity total proposed}
    \end{align}
\end{itemize}
Therefore, the computational complexity of the proposed PDF and CDF
representations grows \emph{linearly} with \(n\) in the precomputation stage
and \emph{quadratically} with \(K\) in the recursive generation of the
mixture weights. Importantly, the quadratic term in \(K\) is independent
of the quadratic-form dimension \(n\).

If the PDF or CDF is evaluated at $N_q$ evaluation points, both the
quantities \(\{S_m\}_{m=1}^{K}\) and the mixture weights
\(\{\beta_k\}_{k=0}^{K}\) are generated only once, since they are
independent of \(q\). Therefore, the corresponding total complexity is
\begin{align}
    \mathcal{C}_{\mathrm{PDF},N_q}
    =
    \mathcal{C}_{\mathrm{CDF},N_q}
    &=
    O\!\left(
        nK+K^2+N_q K
    \right).
    \label{eq: proposed multiple point complexity}
\end{align}

\subsubsection{Complexity of Raphaeli's Method}
\label{sec: complexity Raphaeli method}

For comparison, we analyze the computational complexity of Raphaeli's
PDF and CDF-related series in
\cite[Eqs.~(25) and (32)]{Raphaeli96}. Strictly speaking,
\cite[eq.~(32)]{Raphaeli96} gives the upper-tail probability
\(\Pr\{Q\geq q\}\) for \(q>0\). The conventional CDF is obtained by
taking its complement, which does not alter its computational
complexity.

Let \(N_\lambda\) denote the number of distinct eigenvalues after grouping repeated eigenvalues, and let \(\{\widetilde{\lambda}_k\}_{k=1}^{N_\lambda}\) denote these distinct eigenvalues, with corresponding multiplicities \(\{\varpi_k\}_{k=1}^{N_\lambda}\). Furthermore, define the set of indices associated with the distinct eigenvalues contributing on the selected half-line as
\begin{align}
   \mathcal I_s\triangleq
\begin{cases}
\{k:\widetilde{\lambda}_k>0\}, & q>0,\\
\{k:\widetilde{\lambda}_k<0\}, & q<0,
\end{cases} 
\end{align}
where \(k\in\{1,\ldots,N_\lambda\}\), and let \(N_s\triangleq\operatorname{card}(\mathcal I_s)\). Assume that Raphaeli's infinite series is truncated by retaining \(K_{\mathrm R}+1\) terms for each contributing distinct eigenvalue. Since the series associated with \(\widetilde{\lambda}_k\) begins at derivative order \(\varpi_k-1\), the largest required derivative order is \(D_k\triangleq\varpi_k+K_{\mathrm R}-1\).


The component-wise computational costs are summarized as follows.
\begin{itemize}

    \item \textit{Logarithmic derivatives:}
    For each derivative order \(r\), the expression for
    \(
        \left[
            \log g_k(0,q)
        \right]^{(r)}
    \)
    contains a summation over the distinct eigenvalues. Therefore, evaluating one logarithmic derivative requires
    \(O(N_{\lambda})\) operations. Computing all logarithmic
    derivatives through order \(D_k\) requires
    \(O(N_{\lambda}D_k)\) operations for the \(k\)th contributing
    eigenvalue.

    \item \textit{Recursive derivatives:}
    Raphaeli's recursion for \(g_k^{(r)}(0,q)\) contains \(r\)
    previously computed quantities. Hence, computing the derivative
    of order \(r\) requires \(O(r)\) operations. Generating all
    derivatives through order \(D_k\) consequently requires
    \begin{align}
        \sum_{r=1}^{D_k}O(r)
        &=
        O\!\left(D_k^2\right).
        \label{eq: Raphaeli derivative complexity}
    \end{align}

    \item \textit{Series accumulation:}
    Once the required derivatives have been generated, evaluating and
    accumulating the \(K_{\mathrm{R}}+1\) retained terms requires
    \(O(K_{\mathrm{R}})\) operations for each contributing
    eigenvalue.

\end{itemize}
Consequently, Raphaeli's PDF and upper-tail-probability expressions
have the same computational order:
\begin{align}
    \mathcal{C}_{\mathrm{PDF}}^{(\mathrm{R})}
    =
    \mathcal{C}_{\mathrm{CCDF}}^{(\mathrm{R})}
    &=
    O\!\left(
        \sum_{k\in\mathcal{I}_s}
        \left(
            D_k^2
            +
            N_{\lambda}D_k
            +
            K_{\mathrm{R}}
        \right)
    \right).
    \label{eq: Raphaeli general complexity}
\end{align}
Since the ordinary CDF is obtained by complementation, it follows that $\mathcal{C}_{\mathrm{CDF}}^{(\mathrm{R})}
    =
    \mathcal{C}_{\mathrm{CCDF}}^{(\mathrm{R})}$.
If the eigenvalue multiplicities remain bounded independently of
\(K_{\mathrm{R}}\), then
\(
    D_k=K_{\mathrm{R}}+O(1)
\),
and \eqref{eq: Raphaeli general complexity} simplifies to
\begin{align}
    \mathcal{C}_{\mathrm{PDF}}^{(\mathrm{R})}
    =
    \mathcal{C}_{\mathrm{CDF}}^{(\mathrm{R})}
    &=
    O\!\left(
        N_s K_{\mathrm{R}}^2
        +
        N_s N_\lambda K_{\mathrm{R}}
    \right).
    \label{eq: Raphaeli simplified complexity}
\end{align}

For a definite quadratic form evaluated on its nontrivial support
half-line, all eigenvalues contribute and therefore
\(N_s=N_\lambda\).
When all eigenvalues are distinct, one additionally has
\(
    N_s=N_{\lambda}=n
\)
and
\(
    \varpi_k=1
\).
Thus,
\begin{align}
    \mathcal{C}_{\mathrm{PDF}}^{(\mathrm{R})}
    =
    \mathcal{C}_{\mathrm{CDF}}^{(\mathrm{R})}
    &=
    O\!\left(
        nK_{\mathrm{R}}^2
        +
        n^2K_{\mathrm{R}}
    \right).
    \label{eq: Raphaeli definite complexity}
\end{align}

If the PDF and CDF are evaluated at $N_q$ distinct values of \(q\), the
recursively generated quantities that depend on \(q\) must be
recomputed at each evaluation point. Consequently, the computational complexities for
$N_q$ evaluation points are
\begin{align}
\mathcal{C}_{\mathrm{PDF},N_q}^{(\mathrm{R})}
    =
    \mathcal{C}_{\mathrm{CDF},N_q}^{(\mathrm{R})}
    &=
    O\!\left(
        N_q\left[
            N_s K_{\mathrm{R}}^2
            +
            N_s N_\lambda K_{\mathrm{R}}
        \right]
    \right).
    \label{eq: Raphaeli multiple point complexity}
\end{align}

For a definite quadratic form with distinct eigenvalues, where
\(N_s=N_{\lambda}=n\), this reduces to
\begin{align}
    \mathcal{C}_{\mathrm{PDF},N_q}^{(\mathrm{R})}
    =
    \mathcal{C}_{\mathrm{CDF},N_q}^{(\mathrm{R})}
    &=
    O\!\left[
        N_q
        \left(
            nK_{\mathrm{R}}^2
            +
            n^2K_{\mathrm{R}}
        \right)
    \right].
    \label{eq: Raphaeli multiple point definite complexity}
\end{align}

\subsubsection{Complexity Comparison}
\label{sec: complexity comparison}


For a single evaluation point and equal truncation orders, \(K_{\mathrm{R}}=K\), the two methods
satisfy
\begin{align}
    \mathcal{C}_{\mathrm{PDF}}
    =
    \mathcal{C}_{\mathrm{CDF}}
    &=
    O\!\left(
        nK+K^2
    \right),
    \label{eq: proposed complexity equal truncation}\\
    \mathcal{C}_{\mathrm{PDF}}^{(\mathrm{R})}
    =
    \mathcal{C}_{\mathrm{CDF}}^{(\mathrm{R})}
    &=
    O\!\left(
        nK^2+n^2K
    \right).
    \label{eq: Raphaeli complexity equal truncation}
\end{align}

Thus, under equal truncation orders and the implementation assumptions
adopted above, the proposed method removes an additional multiplicative
dependence on the quadratic-form dimension \(n\). In particular, when
\(K\gg n\), the proposed method scales as \(O(K^2)\), whereas
Raphaeli's formulation scales as \(O(nK^2)\). Conversely, when
\(n\gg K\), the proposed method scales as \(O(nK)\), whereas
Raphaeli's formulation scales as \(O(n^2K)\). Consequently, the
computational advantage of the proposed representation becomes
increasingly significant as either the truncation order or the
quadratic-form dimension grows.


For equal truncation orders \(K_{\mathrm{R}}=K\) and $N_q$ evaluation points, the corresponding computational complexities are
\begin{align}
    \mathcal{C}_{\mathrm{PDF},N_q}
    =
    \mathcal{C}_{\mathrm{CDF},N_q}
    &=
    O\!\left(
        nK+K^2+N_q K
    \right),
    \label{eq: proposed complexity equal truncation M points}\\
    \mathcal{C}_{\mathrm{PDF},N_q}^{(\mathrm{R})}
    =
    \mathcal{C}_{\mathrm{CDF},N_q}^{(\mathrm{R})}
    &=
    O\!\left[
        N_q
        \left(
            nK^2+n^2K
        \right)
    \right].
    \label{eq: Raphaeli complexity equal truncation M points}
\end{align}
Hence, the proposed formulation is particularly advantageous for
multipoint PDF and CDF evaluations. 
Its $q$-independent quantities are computed once and reused, whereas Raphaeli's q-dependent recursive quantities must be recomputed at every evaluation point, making the proposed method particularly advantageous for complete PDF and CDF evaluations.

Finally, it should be emphasized that the truncation orders \(K\) and
\(K_{\mathrm{R}}\) required to achieve the same prescribed accuracy
are not necessarily equal. Therefore, the expressions above characterize
the structural computational complexity of the two methods, whereas
accuracy-matched execution-time comparisons must additionally account
for their different convergence rates.

\subsection{Truncation Error Analysis}
\label{sec: Truncation Error Analysis}

Following the truncation
procedure adopted in Section~\ref{sec: complexity comparison} assumes that the PDF and CDF series are truncated after the term
\(k=K\). Define the corresponding truncated series by
\begin{align}
    f_{Q,K}(q)
    &\triangleq
    \sum_{k=0}^{K}
    \beta_k
    \mathit{f}_{G}\left(\alpha_k,\theta,\left|q\right| \right) 
    \operatorname{u}(q\eta^*),
    &
    q\in\mathbb{R}
    \label{eq: appC truncated PDF}\\
    F_{Q,K}(q)
    &\triangleq
    \frac{1-\sigma}{2}
    +
    \sigma
    \sum_{k=0}^{K}
    \beta_k
    \mathit{F}_{G}\left(\alpha_k,\theta,\left|q\right| \right) 
    \operatorname{u}(q\eta^*),
    &
    q\in\mathbb{R}.
    \label{eq: appC truncated CDF}
\end{align}
The associated truncation errors are
\begin{align}
    \nonumber R_{f,K}(q)
    &\triangleq \left|
    f_Q(q)-f_{Q,K}(q) \right|\\    
    & =
    \sum_{k=K+1}^{\infty}
    \beta_k
    \mathit{f}_{G}\left(\alpha_k,\theta,\left|q\right| \right) 
    \operatorname{u}(q\eta^*), 
    \label{eq: appC PDF remainder}\\
    \nonumber R_{F,K}(q)
    &\triangleq
   \left| F_Q(q)-F_{Q,K}(q) \right|\\   
    & =
     \left|\sigma \right|
    \sum_{k=K+1}^{\infty}
    \beta_k
    \mathit{F}_{G}\left(\alpha_k,\theta,\left|q\right| \right) 
    \operatorname{u}(q\eta^*).
    \label{eq: appC CDF remainder}
\end{align}

As shown in Appendix~\ref{app: truncation error bounds}, the uniform PDF and CDF truncation errors satisfy
\begin{align}
    \sup_{q\in\mathbb{R}}
    R_{f,K}(q)
    &\leq
    \frac{c_{n+K}}{\theta}
    \mathcal{E}_K(\rho),
    &
    1<\rho<R_\beta
    \label{eq: appC PDF error bound}\\
    \sup_{q\in\mathbb{R}}
    R_{F,K}(q)
    &\leq
    \mathcal{E}_K(\rho),
    &
    1<\rho<R_\beta.
    \label{eq: appC CDF uniform bound}
\end{align}
Here, \(R_\beta\) is defined in \eqref{eq: appA radius}, and
\begin{align}
    c_{n+K}
    &\triangleq
    \frac{(n+K)^{n+K}\exp\!\left(-(n+K)\right)}
    {\Gamma(n+K+1)}.
    \label{eq: appC cnK}
\end{align}
Moreover, 
\begin{align}
    \mathcal{E}_K(\rho)
    &\triangleq
    \frac{1}{\rho^{K+1}}
    \prod_{i=1}^{n}
    \frac{r_i}{1-a_i\rho}
    \exp\!\left[
        (\rho-1)
        \sum_{i=1}^{n}
        \frac{\mu_i}{1-a_i\rho}
    \right].
    \label{eq: appC explicit tail bound}
\end{align}

The tightest bound in this family is obtained by minimizing
\(\mathcal{E}_K(\rho)\) over the admissible interval:
\begin{align}
    \mathcal{E}_K^{\star}
    &\triangleq
    \inf_{1<\rho<R_\beta}
    \mathcal{E}_K(\rho).
    \label{eq: appC optimized tail bound}
\end{align}
When numerical optimization is not desired, a simple admissible choice for \(R_\beta<\infty\) is the midpoint
\begin{align}
    \rho_{\mathrm{mid}}
    &\triangleq
        \dfrac{1+R_\beta}{2}.
    \label{eq: appC explicit rho}
\end{align}
When \(R_\beta=\infty\), any finite \(\rho>1\) is admissible.


Our truncation-error bounds in \eqref{eq: appC PDF error bound} and \eqref{eq: appC CDF uniform bound} are of practical importance as they can be used to determine a \emph{sufficient} truncation order $K$ that guarantees a prescribed accuracy for the PDF and CDF, respectively.

\subsection{Closed-Form Approximations}
\label{sec: closed-form approximations}

As will be demonstrated in Section~\ref{sec: Numerical Results}, the
corresponding truncation errors decrease rapidly as the truncation
order \(K\) increases. This rapid convergence, together with the
Gamma-mixture structure of the PDF and CDF expressions in
\eqref{eq: theorem PDF Q} and \eqref{eq: theorem CDF Q}, motivates the
following finite-sum closed-form approximations for the PDF and CDF of
\(Q\).

Define the retained mixture mass as
\begin{align}
    \Omega_K
    &\triangleq
    \sum_{k=0}^{K}\beta_k
    =
    1-T_K,
    \label{eq: closed form retained mass}
\end{align}
\normalsize
where \(T_K\) is the tail mass defined in
\eqref{eq: appC tail mass}. Since \(\beta_0>0\) and
\(\beta_k\geq 0\) for every \(k\in\mathbb{N}_0\), one has
\(
    0<\Omega_K\leq 1.
\)
We then introduce the normalized mixture weights
\begin{align}
    \widetilde{\beta}_{k}^{(K)}
    &\triangleq
    \frac{\beta_k}{\Omega_K},
    \qquad
    k=0,\ldots,K.
    \label{eq: closed form normalized weights}
\end{align}

By construction, these coefficients satisfy $\widetilde{\beta}_{k}^{(K)}
    \geq
    0$ for $k=0,\ldots,K$ and $\sum_{k=0}^{K}
    \widetilde{\beta}_{k}^{(K)}
    =
    1$.
Accordingly, the PDF of \(Q\) can be approximated by the finite
Gamma-mixture representation
\begin{align}
    \widetilde{f}_{Q,K}(q)
    &=
    \sum_{k=0}^{K}
    \widetilde{\beta}_{k}^{(K)}
    f_G\!\left(
        \alpha_k,
        \theta,
        |q|
    \right)
    \operatorname{u}(q\eta^*),
    \qquad
    q\in\mathbb{R}.
    \label{eq: closed form PDF approximation}
\end{align}
Similarly, the CDF of $Q$ can be approximated as
\begin{align}
    \nonumber \widetilde{F}_{Q,K}(q)
    =&
    \frac{1-\sigma}{2}\\
    +& 
    \sigma
    \sum_{k=0}^{K}
    \widetilde{\beta}_{k}^{(K)}
    F_G\!\left(
        \alpha_k,
        \theta,
        |q|
    \right)
    \operatorname{u}(q\eta^*),
    \qquad
    q\in\mathbb{R}.
    \label{eq: closed form CDF approximation}
\end{align}
Since the normalized coefficients in
\eqref{eq: closed form normalized weights} are nonnegative and sum to
unity, \eqref{eq: closed form PDF approximation} represents a valid
PDF for every \(K\in\mathbb{N}_0\). In particular, $\widetilde{f}_{Q,K}(q) \geq0$ and $\int_{-\infty}^{\infty} \widetilde{f}_{Q,K}(q)\,dq= \sum_{k=0}^{K}    \widetilde{\beta}_{k}^{(K)}=1$.
In addition, the sign
factor \(\sigma\) and the additive term \((1-\sigma)/2\) ensure the
correct CDF behavior in the positive- and negative-definite cases. 
In particular, the resulting CDF is nondecreasing, takes values in
\([0,1]\), and satisfies the limiting conditions
\(
\lim_{q\to-\infty}\widetilde{F}_{Q,K}(q)=0\) and
\(\lim_{q\to\infty}\widetilde{F}_{Q,K}(q)=1
\).

\section{Sample Application: Maximal Ratio Combining Over Correlated Rician Channels}
\label{sec: Sample Application}

This section illustrates the direct applicability of the Corollary by
deriving an exact expression for the ABER of an MRC diversity
receiver with an arbitrary number of branches operating over correlated
Rician fading channels.

Consider an MRC diversity receiver with \(P\) branches. The received
signal vector can be expressed as
\begin{align}
    \mathbf{r}
    =
    \mathbf{g}z+\mathbf{w},
    \label{eq: MRC received signal}
\end{align}
where \(z\) is the complex transmitted symbol satisfying
\(\mathbb{E}\{|z|^2\}=E_s\),
\(\mathbf{w}\sim\mathcal{CN}_P(\mathbf{0},N_0\mathbf{I}_P)\)
is the additive noise vector, and
\(\mathbf{g}\in\mathbb{C}^{P\times 1}\) is the correlated Rician
channel vector, modeled as $\mathbf{g}
    \sim
    \mathcal{CN}_P
    \left(
    \overline{\mathbf{g}},
    \mathbf{\Sigma}
    \right)$,
where \(\overline{\mathbf{g}}\in\mathbb{C}^{P\times 1}\) is the
deterministic component and
\(\mathbf{\Sigma}\succ\mathbf{0}\) is the covariance matrix of the
diffuse components.

Under this model, the \(i\)th channel coefficient is marginally Rician
distributed with average power $\Omega_i
    \triangleq
    \mathbb{E}\{|g_i|^2\}
    =
    |\overline{g}_i|^2+\Sigma_{i,i}$,
and Rician factor $\kappa_i
    =
    \frac{|\overline{g}_i|^2}{\Sigma_{i,i}}$.
Thus, the diversity branches are not required to have identical or
unit average powers.

Let the eigendecomposition of the covariance matrix be
\begin{align}
    \mathbf{\Sigma}
    =
    \mathbf{V}
    \mathbf{\Lambda}
    \mathbf{V}^{\dagger},
    \label{eq: Sigma eigendecomposition}
\end{align}
where \(\lambda_i>0\), $\mathbf{\Lambda}
    =
    \operatorname{diag}
    \left(
    \lambda_1,\ldots,\lambda_P
    \right)$, and \(\mathbf{V}\) is unitary. Defining
\(\widetilde{\mathbf{g}}
\triangleq
\mathbf{V}^{\dagger}\overline{\mathbf{g}}\),
the corresponding noncentrality parameters are $\mu_i
    =
    \frac{|\widetilde{g}_i|^2}{\lambda_i}$ with $i=1,\ldots,P$.
Accordingly, the pairs \(\{\lambda_i,\mu_i\}_{i=1}^{P}\) completely
specify the canonical noncentral CGQF representation employed in the
proposed framework.

Assuming perfect channel estimation and symbol synchronization, the
normalized MRC output can be expressed as
\begin{align}
    r_{\mathrm{out}}
    &=
    \frac{\mathbf{g}^{\dagger}\mathbf{r}}
         {\mathbf{g}^{\dagger}\mathbf{g}}
    =
    z+
    \frac{\mathbf{g}^{\dagger}\mathbf{w}}
         {\mathbf{g}^{\dagger}\mathbf{g}}.
    \label{eq: MRC output}
\end{align}
Defining the reference SNR as $\gamma_0 \triangleq \frac{E_s}{N_0}$,
the instantaneous post-combining SNR is
\begin{align}
    \gamma
    =
    \gamma_0\,
    \mathbf{g}^{\dagger}\mathbf{g}.
    \label{eq: Instantaneous SNR def}
\end{align}
Therefore, $Q
    \triangleq
    \frac{\gamma}{\gamma_0}
    =
    \mathbf{g}^{\dagger}\mathbf{I}_P\mathbf{g}$,
which is precisely the positive-definite noncentral CGQF defined in
\eqref{eq: Q def}, with
\(\mathbf{v}=\mathbf{g}\),
\(\mathbf{A}=\mathbf{I}_P\), and \(n=P\).

The Corollary can therefore be applied directly to any integrable
performance measure that depends on the instantaneous SNR. In
particular, for an integrable function \(\psi(\cdot)\),
\begin{align}
    \mathbb{E}\{\psi(\gamma)\}
    =
    \sum_{k=0}^{\infty}
    \beta_k
    \mathbb{E}\{\psi(\gamma_k)\},
    \label{eq: MRC performance Gamma mixture}
\end{align}
where
\(\gamma_k\triangleq\gamma_0G_k\) and
\(G_k\sim\mathrm{Gamma}(\alpha_k,\theta)\).
Since \(\alpha_k=P+k\), each \(\gamma_k\) has the same distribution as
the instantaneous SNR of a single-link Nakagami-\(m\) fading channel
with $m_k= P+k$ and $\overline{\gamma}_k =
    m_k\gamma_0\theta$.
Moreover, since \(\mathbf{A}=\mathbf{I}_P\), the scale parameter
\(\theta\) reduces to the smallest eigenvalue of
\(\mathbf{\Sigma}\), i.e., $\theta
    =
    \lambda_{\min}(\mathbf{\Sigma})>0$.
Thus, the performance of the correlated Rician MRC receiver can be
obtained as weighted sums of the corresponding performance
measures of single-link Nakagami-\(m\) fading channels.



\subsection{Average Bit Error Rate for Square \(M\)-QAM}

Consider coherent square \(M\)-QAM with Gray bit mapping, where
\(\nu\triangleq\log_2\sqrt{M}\). By applying
\eqref{eq: MRC performance Gamma mixture}, the ABER of the considered
MRC receiver is given directly by
\begin{align}
    \overline{P}_b^{(M)}
    =
    \sum_{k=0}^{\infty}
    \beta_k
    \overline{P}_{b}^{(\mathrm{Nak},M)}
    \left(
    m_k,\overline{\gamma}_k
    \right),
    \label{eq: MRC ABER Nakagami mixture}
\end{align}
where
\(\overline{P}_{b}^{(\mathrm{Nak},M)}
(m_k,\overline{\gamma}_k)\)
denotes the exact ABER of Gray-coded square \(M\)-QAM over a single-link
Nakagami-\(m_k\) fading channel, given by
\begin{align}
    \overline{P}_{b}^{(\mathrm{Nak},M)}
    \left(
    m_k,\overline{\gamma}_k
    \right)
    =
    \frac{1}{\sqrt{M}\,\nu}
    \sum_{\ell=1}^{\nu}
    \sum_{r=0}^{N_{\ell}}
    \omega_{r,\ell}^{(M)}
    I_{\xi_{r,k}}
    \left(
    m_k,\frac{1}{2}
    \right),
    \label{eq: Nakagami M-QAM ABER}
\end{align}
where $N_{\ell}
    \triangleq
    \left(1-2^{-\ell}\right)\sqrt{M}-1$, $\xi_{r,k}
    \triangleq
    \left[
    1+
    \frac{
    3(2r+1)^2\overline{\gamma}_k
    }
    {2m_k(M-1)}
    \right]^{-1}$, and 
\begin{align}
    \omega_{r,\ell}^{(M)}
    &\triangleq
    (-1)^{
    \left\lfloor
    \dfrac{r2^{\ell-1}}{\sqrt{M}}
    \right\rfloor}
    \left[
    2^{\ell-1}
    -
    \left\lfloor
    \frac{r2^{\ell-1}}{\sqrt{M}}
    +\frac{1}{2}
    \right\rfloor
    \right].
    \label{eq: QAM weights}
\end{align}
Using
\(\overline{\gamma}_k=m_k\overline{\gamma}\theta\),
the argument \(\xi_{r,k}\) becomes independent of \(k\), namely, $\xi_r
    \triangleq
    \left[
    1+
    \frac{
    3(2r+1)^2\overline{\gamma}\theta
    }
    {2(M-1)}
    \right]^{-1}$.
Consequently, the exact ABER reduces to
\begin{align}
    \overline{P}_b^{(M)}
    =
    \frac{1}{\sqrt{M}\,\nu}
    \sum_{\ell=1}^{\nu}
    \sum_{r=0}^{N_{\ell}}
    \omega_{r,\ell}^{(M)}
    \sum_{k=0}^{\infty}
    \beta_k
    I_{\xi_r}
    \left(
    m_k,\frac{1}{2}
    \right).
    \label{eq: exact MRC M-QAM ABER}
\end{align}
Thus, the exact ABER of the correlated Rician MRC receiver is obtained as weighted sums of the corresponding single-link Nakagami-\(m\) fading results.

\textit{Remark 3:}
Existing studies on MRC over correlated Rician fading provide approximate or asymptotic results for an arbitrary number of branches~\cite{Espinosa19,Ma2005,Mallik2011}, whereas exact results are generally limited to the dual-branch case~\cite{Bithas2007,IlicDelibasic2013}. To the best of our knowledge, \eqref{eq: exact MRC M-QAM ABER} is the first reported exact ABER expression for arbitrary-branch MRC under an arbitrary positive-definite Rician correlation structure. This application also highlights how the proposed Gamma-mixture representation overcomes the limited analytical tractability of Raphaeli's formulation~\cite{Raphaeli96} by reducing further performance measures to weighted sums of single-link Nakagami-\(m\) results. Other modulation schemes and performance metrics (e.g., outage probability) can be treated analogously whenever the corresponding Nakagami-\(m\) error expressions are available in closed form or can be evaluated tractably.

\section{Numerical Results}
\label{sec: Numerical Results}
\begin{table*}[t!]
\centering
\caption{Noncentral CGQF Parameters for $n=17$}
\label{tab: PDF Parameters}
\begin{tabular}{cccccccccccccccccc} 
\toprule
$i$ & $1$  & $2$  & $3$ & $4$ & $5$ & $6$ & $7$ & $8$ & $9$ & $10$ & $11$ & $12$ & $13$ & $14$ & $15$  & $16$  & $17$    \\ \hline \hline  \\
$\lambda_i$  &  $0.60$  & $0.13$  & $0.22$ & $3.82$  & $3.02$  & $0.88$  & $4.02$  & $3.56$  & $0.24$  & $0.42$  & $5.76$  & $3.46$  & $4.22$  & $3.78$  & $0.21$  & $1.32$  & $1.76$   \\  \\
$\mu_i$  &  $0.40$  & $1.20$  & $0.20$ & $0.35$  & $1.05$  & $1.02$  & $0.55$  & $0.30$  & $0.65$  & $1.10$  & $2.78$  & $0.12$ & $0.21$ & $0.40$ & $0.75$ & $0.56$ & $1.77$  \\
\bottomrule
\end{tabular}
\end{table*}

 \begin{figure}[t]
 \centering
 \includegraphics[width=0.85\linewidth]{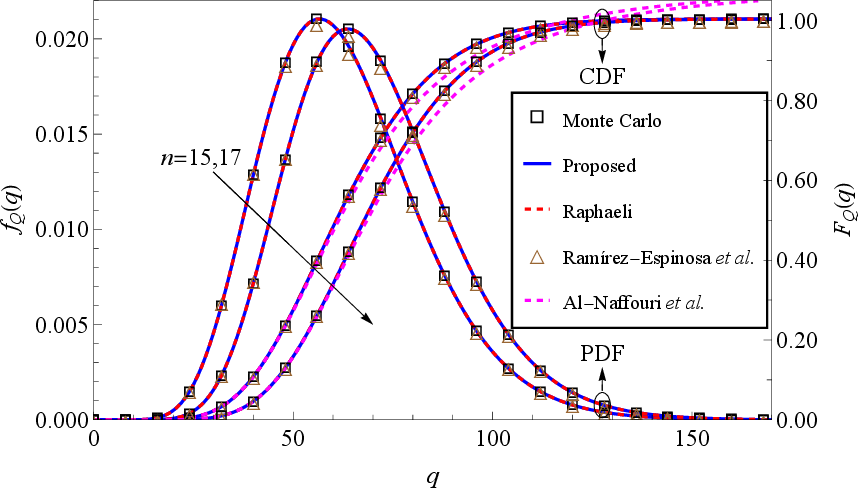}
 \caption{Positive-definite noncentral CGQFs.}
 \label{fig: positive-definite PDF CDF}
 \end{figure}
 \begin{figure}[t]
 \centering
 \includegraphics[width=0.85\linewidth]{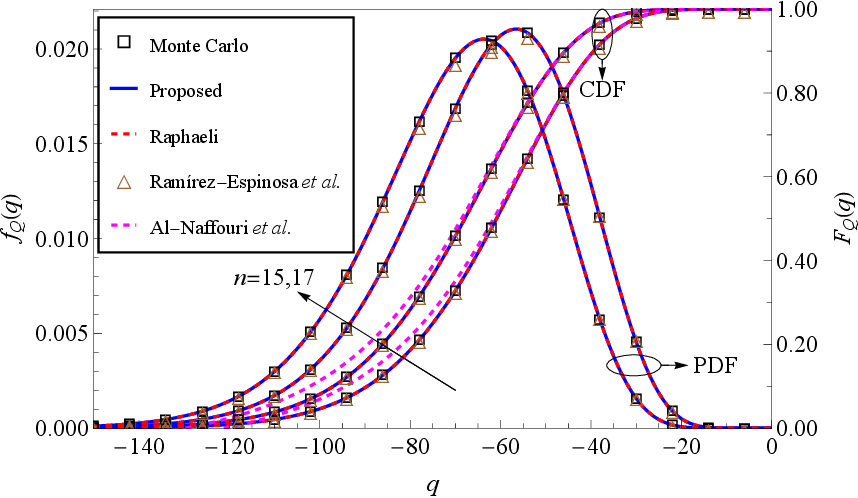}
 \caption{Negative-definite noncentral CGQFs.}
 \label{fig: negative-definite PDF CDF}
 \end{figure}
 \begin{figure}[t]
 \centering
 \includegraphics[width=0.75\linewidth]{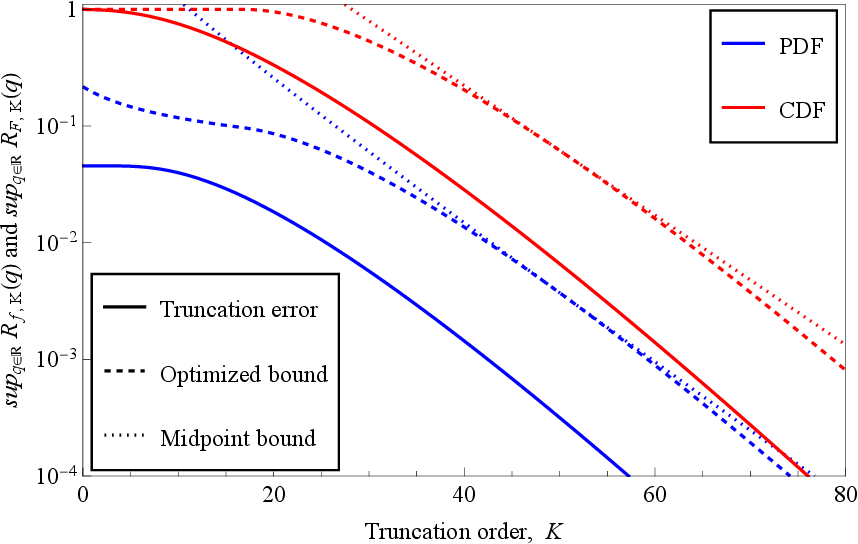}
 \caption{Uniform truncation errors, \(\sup_{q\in\mathbb{R}} R_{f,K}(q)\) and
\(\sup_{q\in\mathbb{R}} R_{F,K}(q)\), and their corresponding uniform bounds as functions of the truncation order, $K$, for $n=10$.}
 \label{fig: Truncation error n10}
 \end{figure}
 \begin{figure}[t]
 \centering
 \includegraphics[width=0.75\linewidth]{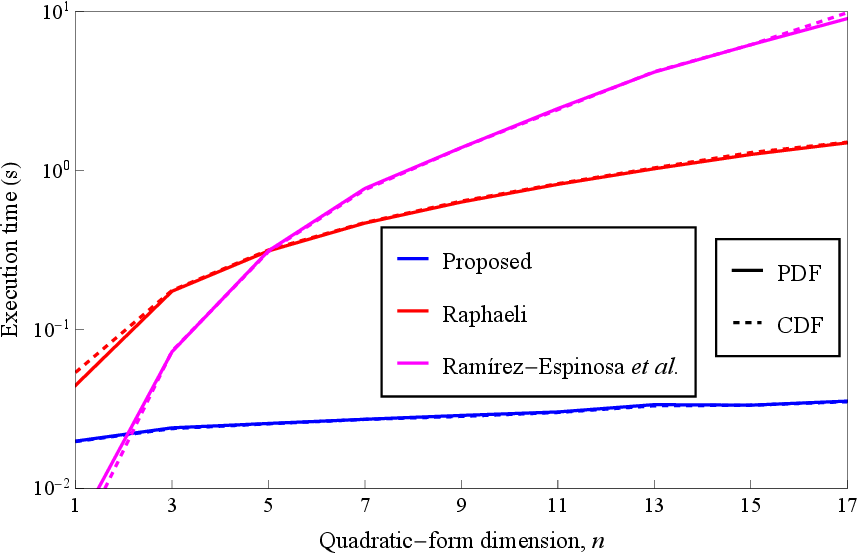}
\caption{Execution time for a single-point evaluation at $q=20$ versus the quadratic-form dimension $n$ in the positive-definite case, subject to an absolute error below $10^{-3}$.}
 \label{fig: Execution time positive}
 \end{figure}
 \begin{figure}[t]
 \centering
 \includegraphics[width=0.75\linewidth]{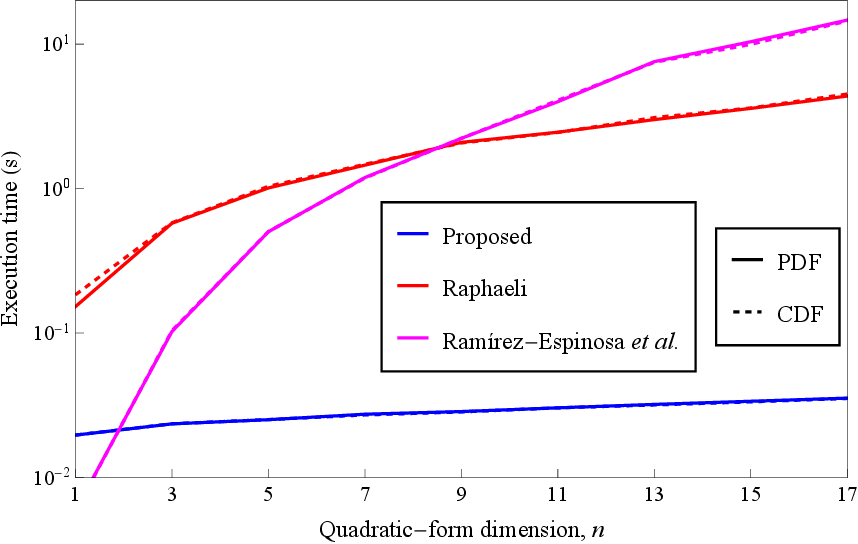}
\caption{Execution time for a single-point evaluation at $q=-20$ versus the quadratic-form dimension $n$ in the negative-definite case, subject to an absolute error below $10^{-3}$.}
 \label{fig: Execution time negative}
 \end{figure}
 \begin{figure}[t]
 \centering
 \includegraphics[width=0.8\linewidth]{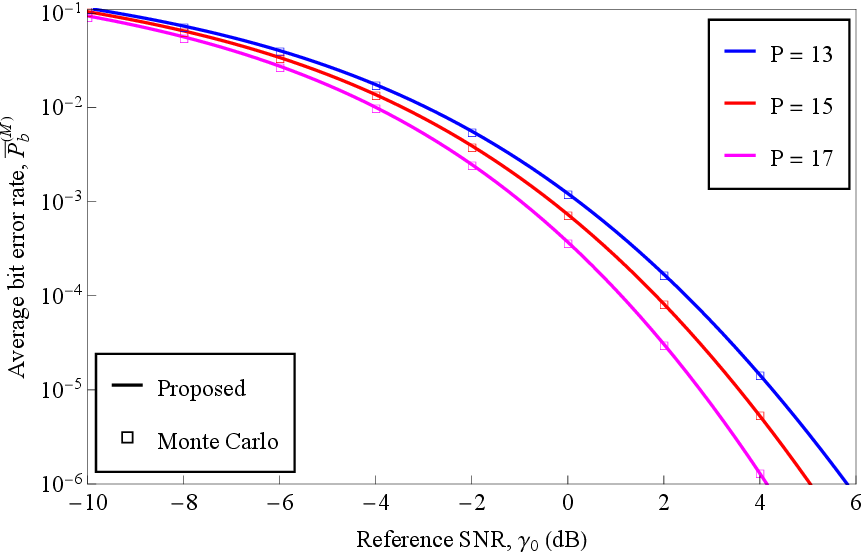}
\caption{ABER versus the reference SNR, \(\gamma_0\), for \(P\in \left\{ 13,15,17\right\} \).}
 \label{fig: ABER}
 \end{figure}

In this section, we numerically corroborate the analytical developments and assess the accuracy, convergence behavior, and computational efficiency of the proposed Gamma-mixture representations. All results are generated using the parameter set in Table~\ref{tab: PDF Parameters}, with the quadratic-form dimension ranging from \(n=1\) to \(n=17\).\footnote{Table~\ref{tab: PDF Parameters} reports the eigenvalue magnitudes for the positive-definite case. For the negative-definite case, the same parameter set is considered with the eigenvalue signs reversed, i.e., \(\lambda_i \mapsto -\lambda_i\), while the noncentrality parameters \(\mu_i\) remain unchanged. For a given dimension $n<17$, the first $n$ parameter pairs in Table~\ref{tab: PDF Parameters} are retained.} The proposed PDF and CDF expressions are compared with the exact series formulation of Raphaeli~\cite[Eqs.~(25) and (32)]{Raphaeli96}, and the approximate formulations of Ramírez-Espinosa \mbox{\emph{et al.}}~\cite[Eqs.~(20) and (24)]{Espinosa19} and Al-Naffouri \mbox{\emph{et al.}}~\cite[Eq.~(41)]{Naffouri16}. Since \mbox{Al-Naffouri} \mbox{\emph{et al.}} derived a CDF-based formulation but no corresponding PDF expression, their results are included only in the CDF comparisons. 

Fig.~\ref{fig: positive-definite PDF CDF} shows the PDF and CDF [Eqs.~\eqref{eq: theorem PDF Q} and \eqref{eq: theorem CDF Q}] for the positive-definite case, whereas Fig.~\ref{fig: negative-definite PDF CDF} shows the corresponding results for the negative-definite case, both for \(n=15\) and \(n=17\).
The proposed expressions in~\eqref{eq: theorem PDF Q} and~\eqref{eq: theorem CDF Q} are evaluated using a truncation order \(K=350\), whereas Raphaeli's series is truncated at \(K_{\mathrm{R}}=375\). For the approximation of Ramírez-Espinosa \mbox{\emph{et al.}}, the accuracy-controlling parameter is set to \(m_{\mathrm{E}}=30\). The Monte Carlo results are obtained from \(10^{7}\) independent realizations.
For both dimensions and both definiteness cases, the proposed PDF and CDF curves, Raphaeli's exact formulation, and the approximation of Ramírez-Espinosa \mbox{\emph{et al.}} are virtually indistinguishable from the Monte Carlo results. This close agreement provides direct numerical validation of the proposed Gamma-mixture representations in~\eqref{eq: theorem PDF Q} and~\eqref{eq: theorem CDF Q}. 
A different behavior is observed for the approximation of \mbox{Al-Naffouri} \mbox{\emph{et al.}} In the positive-definite case, it closely follows the reference CDF in the low-probability region, begins to exhibit a noticeable discrepancy through the transition region, and progressively loses accuracy in the upper tail. Most importantly, its CDF is not constrained to remain within the interval \([0,1]\). 
In the negative-definite case, the behavior is reversed: the largest discrepancy appears in the low-CDF region, whereas the approximation becomes increasingly accurate as the CDF approaches unity. 


Fig.~\ref{fig: Truncation error n10} investigates the convergence of the proposed series by plotting the uniform truncation error bounds for
\(\sup_{q\in\mathbb{R}} R_{f,K}(q)\) [Eq.~\eqref{eq: appC PDF error bound}] and
\(\sup_{q\in\mathbb{R}} R_{F,K}(q)\) [Eq.~\eqref{eq: appC CDF uniform bound}] as functions of the truncation order \(K\), for \(n=10\). \(\sup_{q\in\mathbb{R}} R_{f,K}(q)\) is approximated numerically over a sufficiently dense grid spanning \(0\leq |q|\leq 463\), which covers the numerically relevant range of the distribution, whereas \(\sup_{q\in\mathbb{R}} R_{F,K}(q)\) follows exactly from the remaining mixture mass. The corresponding uniform truncation-error bounds in~\eqref{eq: appC PDF error bound} and~\eqref{eq: appC CDF uniform bound} are shown using both the optimized choice in~\eqref{eq: appC optimized tail bound} and the explicit midpoint choice in~\eqref{eq: appC explicit rho}.
Observe that the actual uniform truncation errors and their corresponding uniform upper bounds decrease monotonically and rapidly with \(K\), confirming the convergence predicted by the truncation-error analysis. 
For instance, the uniform PDF truncation error falls below \(10^{-3}\) at \(K=43\) and below \(10^{-4}\) at \(K=57\). The corresponding uniform CDF truncation error falls below these levels at \(K=62\) and \(K=76\), respectively. 
As expected, the optimized uniform bound is consistently tighter than the explicit midpoint bound. For instance, the optimized PDF bound falls below \(10^{-3}\) at \(K=60\), compared with \(K=43\) for the actual uniform error, while the optimized CDF bound reaches the same level at \(K=79\), compared with \(K=62\) for the actual uniform error. Hence, the bounds are conservative, but remain sufficiently tight to enable practical selection of the truncation order without requiring direct evaluation of the infinite-series remainder.



Figs.~\ref{fig: Execution time positive} and~\ref{fig: Execution time negative} compare the execution times of the proposed method, Raphaeli's formulation, and the approximation of Ramírez-Espinosa \mbox{\emph{et al.}} as the quadratic-form dimension \(n\) increases. For a fair comparison, each method's parameters are independently selected to achieve a pointwise absolute error below \(10^{-3}\). In the positive-definite case, the proposed method uses \(K=190\), Raphaeli's formulation uses \(K_{\mathrm{R}}=220\), and Ramírez-Espinosa \mbox{\emph{et al.}} use \(m_{\mathrm{E}}=25\). For the negative-definite case, these parameters are \(K=190\), \(K_{\mathrm{R}}=375\), and \(m_{\mathrm{E}}=30\), respectively. Single-point PDF and CDF evaluations are performed at \(q=20\) and \(q=-20\) for the positive- and negative-definite cases, respectively.
In the positive-definite case, as \(n\) increases from \(1\) to \(17\), the proposed method's execution time grows modestly from approximately \(1.9\times10^{-2}\) to \(3.4\times10^{-2}\)~s, compared with \(4.6\times10^{-2}\) to \(1.5\)~s for Raphaeli and \(2.3\times10^{-2}\) to \(9.3\)~s for Ramírez-Espinosa \mbox{\emph{et al.}} Consequently, at \(n=17\), the proposed method is approximately \(43\) and \(270\) times faster, respectively.
The computational advantage is even greater in the negative-definite case, where the proposed method's execution time increases only from approximately \(2.0\times10^{-2}\) to \(3.5\times10^{-2}\)~s, compared with \(1.6\times10^{-1}\) to \(4.4\)~s for Raphaeli and \(2.0\times10^{-2}\) to \(14.8\)~s for Ramírez-Espinosa \mbox{\emph{et al.}} At \(n=17\), this corresponds to speedups of approximately \(125\) and more than \(420\), respectively.
Overall, the proposed Gamma-mixture formulation exhibits substantially better execution-time scaling over \(1\leq n\leq17\), with its computational advantage becoming increasingly significant as \(n\) grows. Since these comparisons involve only single-point evaluations, even greater gains are expected for multipoint PDF and CDF evaluations, where the proposed method reuses its evaluation-independent mixture weights.

Finally, Fig.~\ref{fig: ABER} illustrates the ABER [Eq.~\eqref{eq: exact MRC M-QAM ABER}] as a function of the reference SNR for \(P\in\{13,15,17\}\), using the parameter sets in Table~\ref{tab: PDF Parameters} and Gray-coded \(16\)-QAM.
The analytical results are in excellent agreement with the Monte Carlo simulations and, as expected, the system performance improves as the number of diversity branches increases.



\section{Conclusions}
\label{sec: Conclusions}
In this paper, we derived novel exact Gamma-mixture representations for definite noncentral CGQFs, yielding analytically tractable and computationally efficient formulations. As a direct application of the proposed framework, we obtained an exact  ABER expression for MRC receivers with an arbitrary number of diversity branches under an arbitrary positive-definite Rician correlation structure. The absolute and uniform convergence of the resulting PDF and CDF series, together with explicit uniform truncation-error bounds, was established, providing rigorous control over the numerical accuracy. Moreover, the Gamma-mixture structure enables finite-sum approximations that, by construction, remain valid probability distributions for any truncation order. Numerical results confirm the rapid convergence of the proposed series and demonstrate substantial computational gains over the formulations of Raphaeli and Ramírez-Espinosa \mbox{\emph{et al.}}

\begin{appendices}

\section{Proof of the Theorem}
\label{app: proof theorem}

Define $X_i \triangleq \eta_i Y_i$, $i\in\{1,\ldots,n\}$.
Since \(Y_i\sim\chi_2^2(2\mu_i)\), the random variable $X_i$ was shown to follow a Poisson-weighted Gamma mixture, whose PDF is given by~\cite{Johnson1994} 
\begin{align}
    f_{X_i}(x)
    =
    \sum_{k=0}^{\infty}
    \phi_{k,i}
    f_G\!\left(1+k,2|\eta_i|,|x|\right)
    \operatorname{u}(x\eta_i),
    \qquad x\in\mathbb{R},
    \label{eq: appA marginal mixture}
\end{align}
where $f_G(\cdot,\cdot,\cdot)$ denotes the PDF of a Gamma distribution, as defined in~\eqref{eq: PDF G Gamma marginal}. Moreover, $\left\{ \phi_{k,i} \right\}$ are the Poisson mixture weights given by $\phi_{k,i}
    =
    \frac{\mu_i^k}{k!}\exp(-\mu_i)$.
Taking the bilateral Laplace transform of \(X_i\), we obtain
\begin{align}
    \Laplace\{f_{X_i}\}(s)
    &=
    \frac{\exp(-\mu_i)}
    {1+2\eta_i s}
    \exp\!\left(
        \frac{\mu_i}{1+2\eta_i s}
    \right),
    \label{eq: two exp}
\end{align}
where $s \in \mathbb{C}$.
Using the Taylor expansion of the second exponential term in \eqref{eq: two exp}, the bilateral Laplace transform can be written as
\begin{align}
\Laplace \{f_{X_i}\}(s)
&=
\sum_{k=0}^{\infty}
\phi_{k,i}
\left(1+2\eta_i s\right)^{-1-k},
\label{eq: appA marginal LT}
\end{align}
with the corresponding region of convergence (ROC) determined by
\begin{align}
    \Re(1+2\eta_i s)>0
    \Longleftrightarrow
    \begin{cases}
        \Re(s)>-\dfrac{1}{2\eta_i},
        & \eta_i>0\\[2mm]
        \Re(s)<\dfrac{1}{2|\eta_i|},
        & \eta_i<0.
    \end{cases}
    \label{eq: appA marginal ROC}
\end{align}

By the definition of \(\eta^*\) in~\eqref{eq: eta cases}, it follows that $0<r_i\leq1$ and $0\leq a_i<1$.

Define $b_i
    \triangleq
    |\eta_i|$, $b_*
    \triangleq
    |\eta^*|
    =
    \min_i \left\{ b_i \right\}$, $a_{\max} 
    \triangleq
    \max_{i}a_i$, and $b_{\max}
    \triangleq
    \max_i \left\{ b_i \right\}$.
Since all coefficients \(\eta_i\) have the same sign in the definite case, it follows from \eqref{eq: appA marginal ROC} that the intersection of the individual regions of convergence is
\begin{align}
    \sigma\Re(s)
    >
    -\frac{1}{2b_{\max}}.
    \label{eq: appA common ROC}
\end{align}
Hence, \eqref{eq: appA common ROC} provides a common region of
convergence for the bilateral Laplace transforms of all the random
variables \(X_i\). Within this region,
\begin{align}
    \Re\!\left(1+2\eta^*s\right)
    &=
    1+2\sigma b_*\Re(s)>
    1-\frac{b_*}{b_{\max}}
   \geq
    1-\frac{b_*}{b_i}
    =
    a_i, 
    \label{eq: ineq ai}
\end{align}
where the second inequality follows from \(b_i\leq b_{\max}\).
Since $0\leq a_i<1$ and $\Re\!\left(1+2\eta^*s\right)>a_i\geq 0$, we have
\begin{align}
    \left|a_i t\right|
    &=
    \frac{a_i}
    {\left|1+2\eta^*s\right|}
    \leq
    \frac{a_i}
    {\Re\!\left(1+2\eta^*s\right)}
    <1.
    \label{eq: appA convergence condition}
\end{align}
Therefore, the power-series expansions employed below converge
absolutely throughout the common ROC.

Defining $t 
    \triangleq
    \frac{1}{1+2\eta^*s}$, together with \(a_i=1-r_i\), we obtain $1+2\eta_i s=
    \frac{1-a_i t}
    {r_i t}$.
Substituting this result
into~\eqref{eq: appA marginal LT} and collecting the terms
that depend on \(t\), the marginal bilateral Laplace transform
can be rewritten as $\Laplace\{f_{X_i}\}(s)
    =
    r_i t
    \frac{\exp(-\mu_i)}
    {1-a_i t}
    \exp\!\left(
        \frac{\mu_i r_i t}
        {1-a_i t}
    \right)$.
Since the random variables \(\{X_i\}_{i=1}^{n}\) are mutually
independent, the bilateral Laplace transform of
\(Q=\sum_{i=1}^{n}X_i\) is given by the product of the corresponding
marginal transforms. 
Therefore,
\begin{align}
    \Laplace\{f_Q\}(s)
    &=
    \prod_{i=1}^{n}
    \Laplace\{f_{X_i}\}(s)
    =
    t^n B\!\left(t\right),
    \label{eq: appA LT Q product}
\end{align}
where 
\begin{align}
    B\!\left(t\right)
    &\triangleq
    \prod_{i=1}^{n}
    \frac{r_i}{1-a_i t}
    \exp\!\left(
        \frac{\mu_i\bigl(t-1\bigr)}
        {1-a_i t}
    \right)
    =
    \sum_{k=0}^{\infty}
    \beta_k t^k.
    \label{eq: appA beta generating function}
\end{align}
The second equality defines \(\{\beta_k\}_{k=0}^{\infty}\) as the Taylor coefficients of \(B(t)\) about \(t=0\).

From the definition in \eqref{eq: appA beta generating function}, the possible finite singularities of \(B(t)\) occur at \( t=\frac{1}{a_i} \), for  \(a_i>0\). Since \(0\leq a_i\leq a_{\max}\), the singularity closest to the origin is located at \(t=1/a_{\max}\) whenever \(a_{\max}>0\). Therefore, \(B(t)\) is analytic in the open disk \( |t|<R_\beta, \) where \begin{align} R_\beta &\triangleq \begin{cases} \dfrac{1}{a_{\max}}, & a_{\max}>0,\\[2mm] \infty, & a_{\max}=0. \end{cases} \label{eq: appA radius} \end{align} When \(a_{\max}=0\), all \(a_i=0\), and hence \(B(t)\) is an entire function. Consequently, the Taylor series of \(B(t)\) about \(t=0\) converges absolutely for every \(t\) satisfying \( |t|<R_\beta\).

Condition~\eqref{eq: appA convergence condition} guarantees that, for every
\(s\) in the common ROC, the corresponding value
\(
    t=\frac{1}{1+2\eta^*s}
\)
lies inside the disk of analyticity of \(B(t)\).
Therefore, the Taylor expansion of \(B(t)\) in
\eqref{eq: appA beta generating function} can be evaluated at this value of
\(t\). Substituting the resulting expansion into
\eqref{eq: appA LT Q product} yields
\begin{align}
    \Laplace\{f_Q\}(s)
    &=
    t^n
    \sum_{k=0}^{\infty}
    \beta_k t^k
    =
    \sum_{k=0}^{\infty}
    \beta_k
    \left(1+2\eta^*s\right)^{-n-k}.
    \label{eq: appA LT Q series}
\end{align}

Setting \(t=0\) in~\eqref{eq: appA beta generating function} gives the
zeroth-order coefficient,
\begin{align}
    \beta_0
    &=
    B(0)
   =
    \left(
        \prod_{i=1}^{n}
        r_i
    \right)
    \exp\!\left(
        -\sum_{i=1}^{n}\mu_i
    \right).
    \label{eq: appA beta zero}
\end{align}

To derive the remaining coefficients, we logarithmically differentiate
\(B(t)\). For \(|t|<R_\beta\), this gives
\begin{align}
    \frac{tB'(t)}{B(t)}
    &=
    \sum_{i=1}^{n}
    \left[
        \frac{a_i t}{1-a_i t}
        +
        \frac{\mu_i r_i t}
        {(1-a_i t)^2}
    \right].
    \label{eq: appA logarithmic derivative intermediate}
\end{align}
Here, the prime symbol \(('\)) denotes differentiation with respect to \(t\).
Using the power-series identities
\begin{align}
    \frac{a_i t}{1-a_i t}
    =&
    \sum_{m=1}^{\infty}
    a_i^m t^m \quad \text{and} \quad 
    \frac{t}{(1-a_i t)^2}
    =
    \sum_{m=1}^{\infty}
    m a_i^{m-1}t^m,
\end{align}
we obtain
\begin{align}
    \frac{tB'(t)}{B(t)}
    &=
    \sum_{m=1}^{\infty}
    \mathcal{S}_m t^m,
    \label{eq: appA logarithmic derivative}
\end{align}
where
\begin{align}
    \mathcal{S}_m
    &\triangleq
    \sum_{i=1}^{n}
    a_i^{m-1}
    \left(
        a_i+m\mu_i r_i
    \right),
    \qquad m\geq 1.
    \label{eq: appA Sm}
\end{align}

Multiplying both sides of
\eqref{eq: appA logarithmic derivative} by \(B(t)\) gives
\begin{align}
    tB'(t)
    &=
    B(t)
    \sum_{m=1}^{\infty}
    \mathcal{S}_m t^m.
    \label{eq: appA coefficient identity}
\end{align}

Since \(
    tB'(t)
    =
    \sum_{k=1}^{\infty}
    k\beta_k t^k\)
and \(
    B(t)
    \sum_{m=1}^{\infty}
    \mathcal{S}_m t^m
    =
    \sum_{k=1}^{\infty}
    \left(
        \sum_{m=1}^{k}
        \beta_{k-m}\mathcal{S}_m
    \right)t^k\), 
equating the coefficients of \(t^k\) yields~\cite{Garcia24}
\begin{align}
    \beta_k
    &=
    \frac{1}{k}
    \sum_{m=1}^{k}
    \beta_{k-m}\mathcal{S}_m,
    \qquad k\geq 1.
    \label{eq: appA beta recursion compact}
\end{align}

Finally, substituting \eqref{eq: appA Sm} into \eqref{eq: appA beta recursion compact} and using the initial coefficient in \eqref{eq: appA beta zero}, we recover the recursion given in \eqref{eq: Coefficients beta final}.

Since \(0\leq a_i<1\), \(0<r_i\leq 1\), and \(\mu_i\geq 0\), every term in \eqref{eq: appA Sm} is nonnegative. Consequently, $\mathcal{S}_m \geq 0$ for $m\geq 1$. 
Moreover, \eqref{eq: appA beta zero} implies that \(\beta_0>0\). It then follows by induction from the recursion in
\eqref{eq: appA beta recursion compact} that
\begin{align}
    \beta_k
    &\geq
    0,
    \qquad
    k\in\mathbb{N}_0.
    \label{eq: appA beta nonnegative}
\end{align}

Furthermore, because \(a_{\max}<1\), the radius of convergence satisfies
\(R_\beta>1\), and therefore the Taylor series of \(B(t)\) converges
absolutely at \(t=1\). Evaluating \eqref{eq: appA beta generating function}
at \(t=1\) and using the definition \(a_i \triangleq 1-r_i\), we obtain
\begin{align}
    \sum_{k=0}^{\infty}\beta_k
    &=
    B(1) =
    \prod_{i=1}^{n}
    \frac{r_i}{1-a_i}
    \exp\!\left(
        \frac{\mu_i(1-1)}
        {1-a_i}
    \right)=
    1.
    \label{eq: appA beta normalization}
\end{align}
Hence, the sequence \(\{\beta_k\}_{k=0}^{\infty}\) consists of
nonnegative coefficients that sum to unity and therefore constitutes a
valid sequence of mixture weights.


For \(\alpha_k>0\), \(\theta>0\), and
\(\sigma\in\{-1,1\}\), the following bilateral Laplace-transform
pair holds:
\begin{align}
    \Laplace
    \left\{
        f_G\!\left(\alpha_k,\theta,|q|\right)
        \operatorname{u}(\sigma q)
    \right\}(s)
    &=
    \left(1+\sigma\theta s\right)^{-\alpha_k},
    \label{eq: appA unified transform pair}
\end{align}
\normalsize
with region of convergence $\sigma\Re(s)
    >
    -\frac{1}{\theta}$.

Since
\(
    \eta^*
    =
    \sigma|\eta^*|
\)
and
\(
    \theta
    =
    2|\eta^*|
\),
it follows that
\(
    1+2\eta^*s
    =
    1+\sigma\theta s
\).
Therefore, each factor
\(
    \left(1+2\eta^*s\right)^{-\alpha_k}
\)
in \eqref{eq: appA LT Q series} coincides with the bilateral Laplace
transform in \eqref{eq: appA unified transform pair}. Consequently,
taking the inverse bilateral Laplace transform term by term yields a
Gamma density with shape parameter \(\alpha_k\) and scale parameter
\(\theta\), supported on the half-line determined by the sign of
\(\eta^*\).

For real \(s\) in the common region of convergence (ROC), the bilateral Laplace-transform kernel \(e^{-s x}\) and all Gamma-mixture terms are nonnegative. Tonelli's theorem therefore permits the interchange of the infinite summation and the bilateral Laplace integral, yielding a valid transform identity for all real \(s\) in the common ROC.
Both sides are analytic functions of \(s\) in the interior of the ROC. Hence, by the identity theorem~\cite{Kreyszig10}, the equality extends to all complex \(s\) in the same region. This extension permits term-by-term inversion, and uniqueness of the bilateral Laplace transform establishes that the PDF of \(Q\) is given by \eqref{eq: theorem PDF Q}.

Moreover, since the summands in \eqref{eq: theorem PDF Q} are
nonnegative, Tonelli's theorem permits the interchange of the infinite
summation and integration. Because \(\beta_k\geq 0\) for every
\(k\in\mathbb{N}_0\) and
\(
    \sum_{k=0}^{\infty}\beta_k=1,
\)
term-by-term integration of the PDF series yields the CDF
representation in \eqref{eq: theorem CDF Q}. For the
negative-definite case, the upper incomplete-Gamma terms are converted
to the corresponding lower incomplete-Gamma form by using $\Upsilon(a,x)+\Gamma(a,x)
    =
    \Gamma(a)$~\cite{Olver10}.
This completes the proof.

\section{Absolute and Uniform Convergence \\ of the PDF and CDF of $Q$}
\label{app: series convergence}

Define the component Gamma PDF and CDF as
\begin{align}
    g_k(x)
    &\triangleq
    f_G\!\left(\alpha_k,\theta,x\right),
    \qquad
    x\geq 0
    \label{eq: appB component PDF}\\
    G_k(x)
    &\triangleq
    F_G\!\left(\alpha_k,\theta,x\right),
    \qquad
    x\geq 0,
    \label{eq: appB component CDF}
\end{align}
and let $ d_k
    \triangleq
    \alpha_k-1
    =
    n+k-1 \in \mathbb{N}_0$.

The maximum value of \(g_k(x)\) over \(x\geq 0\) is
given by~\cite{Johnson1994}
\begin{align}
    \sup_{x\geq 0} g_k(x)
    &=
    \frac{c_{d_k}}{\theta},
    \label{eq: appB gamma maximum}
\end{align}
where
\begin{align}
    c_0
    &\triangleq
    1,
    \label{eq: appB c0}\\
    c_m
    &\triangleq
    \frac{m^m\exp(-m)}
    {\Gamma(m+1)},
    \qquad
    m\geq 1.
    \label{eq: appB cm}
\end{align}
For \(d_k=0\), \(g_k(x)\) reduces to an exponential
density and attains its maximum at \(x=0\).

Using the standard bound
\(
    \Gamma(d_k+1)
    =
    d_k!
    \geq
    \left(\frac{d_k}{e}\right)^{d_k}
\)
for \(d_k\geq 1\)~\cite{prudnikov92Vol2},
we obtain \(0<c_{d_k}\leq 1\) for \(d_k\geq 1\).
Together with \(c_0=1\), this implies the uniform bound
\begin{align}
    0
    \leq
    g_k(x)
    &\leq
    \frac{1}{\theta},
    \qquad
    x\geq 0,
    \quad
    k\in\mathbb{N}_0.
    \label{eq: appB gamma PDF bound}
\end{align}

Using \eqref{eq: appA beta nonnegative},
\eqref{eq: appA beta normalization}, the bound in
\eqref{eq: appB gamma PDF bound}, and
\(0\leq \operatorname{u}(q\eta^*)\leq 1\), we obtain $\left|
        \beta_k
        g_k(|q|)
        \operatorname{u}(q\eta^*)
    \right|
    \leq
    \frac{\beta_k}{\theta}$ for $q\in\mathbb{R}$.
Moreover, $\sum_{k=0}^{\infty}
    \frac{\beta_k}{\theta}
    =
    \frac{1}{\theta}
    \sum_{k=0}^{\infty}\beta_k
    =
    \frac{1}{\theta}
    <\infty$.
Therefore, the Weierstrass \(M\)-test guarantees that the PDF series
converges absolutely and uniformly on \(\mathbb{R}\).

Similarly, since each \(G_k(x)\) is a Gamma CDF, $0
    \leq
    G_k(x)
    \leq
    1$, for $x\geq 0$.
Hence, $\left|
        \beta_k
        G_k(|q|)
        \operatorname{u}(q\eta^*)
    \right|
    \leq
    \beta_k$, for $q\in\mathbb{R}$.
Since $\sum_{k=0}^{\infty}\beta_k= 1$,
the Weierstrass \(M\)-test likewise establishes the absolute and
uniform convergence of the CDF series on \(\mathbb{R}\). The finite
additive constant and the sign factor appearing in
\eqref{eq: theorem CDF Q} do not affect this convergence result.

Consequently, both the PDF and CDF series converge absolutely and
uniformly over their respective domains. 
This concludes the proof.

\section{Proof of Corollary~\ref{cor: expectation Q}}
\label{app: proof expectation Q}

By the definition of expectation, we have $\mathbb{E}\!\left\{\psi(Q)\right\}
=
\int_{-\infty}^{\infty}
\psi(q)f_Q(q)\,\mathrm{d}q$.
Thus, substituting the PDF in
\eqref{eq: theorem PDF Q}, we obtain
\begin{equation}
\mathbb{E}\!\left\{\psi(Q)\right\}
=
\int_{-\infty}^{\infty}
\psi(q)
\sum_{k=0}^{\infty}
\beta_k
f_G(\alpha_k,\theta,|q|)
u(q\eta^{*})
\,\mathrm{d}q.
\label{eq: expectation Q substituted appendix}
\end{equation}

To justify interchanging the infinite summation and the
integral, first apply Tonelli's theorem to the corresponding
nonnegative integrand involving \(|\psi(q)|\). Since
\(\beta_k\geq 0\) \(\forall k\), we obtain
\begin{align}
\mathbb{E}\!\left\{|\psi(Q)|\right\}
&=
\int_{-\infty}^{\infty}
|\psi(q)|
\sum_{k=0}^{\infty}
\beta_k
f_G(\alpha_k,\theta,|q|)
u(q\eta^{*})
\,\mathrm{d}q
\nonumber\\
&=
\sum_{k=0}^{\infty}
\beta_k
\int_{-\infty}^{\infty}
|\psi(q)|
f_G(\alpha_k,\theta,|q|)
u(q\eta^{*})
\,\mathrm{d}q.
\label{eq: absolute expectation interchange}
\end{align}
The left-hand side of
\eqref{eq: absolute expectation interchange} is finite by
hypothesis. Therefore, Fubini's theorem permits the
interchange of the summation and integration in
\eqref{eq: expectation Q substituted appendix}, yielding
\begin{equation}
\mathbb{E}\!\left\{\psi(Q)\right\}
=
\sum_{k=0}^{\infty}
\beta_k
\int_{-\infty}^{\infty}
\psi(q)
f_G(\alpha_k,\theta,|q|)
u(q\eta^{*})
\,\mathrm{d}q.
\label{eq: expectation Q interchanged appendix}
\end{equation}

Because \(\eta^{*}\) and \(\sigma\) have the same sign, the
unit-step function restricts the integral to the half-line
\(
\mathcal{H}_{\sigma}
\triangleq
\{q\in\mathbb{R}:\sigma q\geq 0\}
\).
Now, applying the change of variables
\(
q=\sigma x\), for \(
x\geq 0\),
to each integral in
\eqref{eq: expectation Q interchanged appendix} gives
\begin{align}
\nonumber & \int_{-\infty}^{\infty}
\psi(q)
f_G(\alpha_k,\theta,|q|)
u(q\eta^{*})
\,\mathrm{d}q\\
& \qquad =
\int_{0}^{\infty}
\psi(\sigma x)
f_G(\alpha_k,\theta,x)
\,\mathrm{d}x  =
\mathbb{E}\!\left\{\psi(\sigma G_k)\right\}.
\label{eq: component expectation appendix}
\end{align}
Substituting \eqref{eq: component expectation appendix} into
\eqref{eq: expectation Q interchanged appendix} yields
\eqref{eq: expectation Q gamma mixture}.

It remains to establish the absolute convergence of the series.
Using the inequality
\(
|\mathbb{E}\{X\}|
\leq
\mathbb{E}\{|X|\}
\),
together with the nonnegativity of the mixture weights and
Tonelli's theorem, we obtain
\begin{align}
\sum_{k=0}^{\infty} &
\left|
\beta_k
\mathbb{E}\!\left\{\psi(\sigma G_k)\right\}
\right|
=
\sum_{k=0}^{\infty}
\beta_k
\left|
\mathbb{E}\!\left\{\psi(\sigma G_k)\right\}
\right|
\nonumber\\
&\leq
\sum_{k=0}^{\infty}
\beta_k
\mathbb{E}\!\left\{|\psi(\sigma G_k)|\right\}
=
\mathbb{E}\!\left\{|\psi(Q)|\right\}
<\infty.
\label{eq: absolute convergence expectation appendix}
\end{align}
Hence, the series in
\eqref{eq: expectation Q gamma mixture} converges absolutely.
If \(\psi\) is nonnegative, Tonelli's theorem applies directly,
and the identity remains valid without requiring the
expectation to be finite, with both sides interpreted in the
extended nonnegative real numbers. This completes the proof.

\section{Closed-Form Truncation-Error Bounds}
\label{app: truncation error bounds}

Assume that the PDF and CDF series are truncated after the term
\(k=K\), and define the remaining mixture mass as
\begin{align}
    T_K
    &\triangleq
    \sum_{k=K+1}^{\infty}\beta_k.
    \label{eq: appC tail mass}
\end{align}
Let \(\rho\) be any real number satisfying $1<\rho<R_\beta$.
Since \(\beta_k\geq 0\) for all \(k\in\mathbb{N}_0\) and
\(\rho^k\geq \rho^{K+1}\) for every \(k\geq K+1\), we have
\begin{align}
    \rho^{K+1}T_K
    &=
    \rho^{K+1}
    \sum_{k=K+1}^{\infty}\beta_k
    \nonumber\\
    &\leq
    \sum_{k=K+1}^{\infty}\beta_k\rho^k
    \leq
    \sum_{k=0}^{\infty}\beta_k\rho^k
    =
    B(\rho).
\end{align}
Consequently,
\begin{align}
    T_K
    &\leq
    \frac{B(\rho)}{\rho^{K+1}}
    \triangleq
    \mathcal{E}_K(\rho).
    \label{eq: appC tail bound}
\end{align}
Using
\eqref{eq: appA beta generating function}, the bound can be written
explicitly as in \eqref{eq: appC explicit tail bound}.


The sequence \(\{c_m\}_{m\geq 1}\) in
\eqref{eq: appB cm} is strictly decreasing since
\(
    \frac{c_{m+1}}{c_m}
    =
    \frac{1}{e}
    \left(1+\frac{1}{m}\right)^m
    <1
\)
for \(m\geq 1\).
For every omitted term \(k\geq K+1\), we have
\(d_k=n+k-1\geq n+K\).
Therefore, by the monotonicity of \(c_m\) and
\eqref{eq: appB gamma maximum},
\begin{align}
    \sup_{x\geq 0} g_k(x)
    &=
    \frac{c_{d_k}}{\theta}
    \leq
    \frac{c_{n+K}}{\theta}.
    \label{eq: appC omitted gamma bound}
\end{align}

Using \eqref{eq: appC omitted gamma bound},
\(\beta_k\geq 0\),
\(0\leq\operatorname{u}(q\eta^*)\leq 1\),
and the definition of \(T_K\), it follows that
\begin{align}
    R_{f,K}(q)
    &=
    \sum_{k=K+1}^{\infty}
    \beta_k
    g_k(|q|)
    \operatorname{u}(q\eta^*)
    \nonumber\\
    &\leq
    \frac{c_{n+K}}{\theta}
    \sum_{k=K+1}^{\infty}\beta_k
    =
    \frac{c_{n+K}}{\theta}T_K.
\end{align}
Combining this result with
\eqref{eq: appC tail bound}, we obtain the uniform PDF
truncation-error bound given in
\eqref{eq: appC PDF error bound}.



Since $0\leq G_k(x)\leq 1$ for $x\geq 0$,
and \(|\sigma|=1\), it follows that
\begin{align}
    R_{F,K}(q)
    &\leq
    \sum_{k=K+1}^{\infty}\beta_k
    =
    T_K.
\end{align}
Combining this inequality with \eqref{eq: appC tail bound} yields the uniform CDF truncation-error bound in \eqref{eq: appC CDF uniform bound}. This completes the proof.

\end{appendices}

\bibliographystyle{IEEEtran}
\bibliography{Bibliography}

\end{document}